\documentclass[final,3p,times]{elsarticle}

\usepackage{amssymb}
\usepackage{lipsum}
\usepackage[utf8]{inputenc}
\usepackage[english]{babel}
\usepackage{hyperref}
\usepackage{graphicx}
\usepackage{amsmath}
\usepackage{mathtools}
\usepackage{color}

\usepackage{dcolumn}
\usepackage{bm}
\usepackage{color}

\usepackage{tabularx}
\usepackage{multirow}
\usepackage{selinput}
\usepackage{array}
\usepackage{booktabs, adjustbox}
\usepackage{amsthm}
\journal{Chaos, Solitons $\&$ Fractals}

\begin{document}

\begin{frontmatter}



\title{Structural Asymmetry as a Fraud Signature: Detecting Collusion with Heron’s Information Coefficient}


\author[label1]{Allana Tavares Bastos}

\author[label2]{Tiago Alves Schieber}

\author[label3]{Renato Hadad}

\author[label4]{Laura Carpi}

\author[label1]{Martín Gómez Ravetti}

\affiliation[label1]{organization={Departamento de Ciência da Computação, Universidade Federal de Minas Gerais},
        city={Belo Horizonte},
        state={MG},
        country={Brazil}
}

\affiliation[label2]{organization={Departamento de Ciências Administrativas, Universidade Federal de Minas Gerais},
        city={Belo Horizonte},
        state={MG},
        country={Brazil}
}

\affiliation[label3]{organization={Departamento de Engenharia de Produção, Pontifícia Universidade Católica de Minas Gerais},
        city={Belo Horizonte},
        state={MG},
        country={Brazil}
}

\affiliation[label4]{organization={FutureTech Ltda.},
        city={Belo Horizonte},
        state={MG},
        country={Brazil}
}

\begin{abstract}
Fraud in public procurement remains a persistent challenge, especially in large, decentralized systems like Brazil’s Unified Health System. We introduce Heron’s Information Coefficient (HIC), a geometric measure that quantifies how subgraphs deviate from the global structure of a network. Applied to over eight years of Brazilian bidding data for medical supplies, this measure highlights possible collusive patterns that standard indicators may overlook. Unlike conventional robustness metrics, the Heron coefficient focuses on the interaction between active and inactive subgraphs, revealing structural shifts that may signal coordinated behavior, such as cartel formation. Synthetic experiments support these findings, demonstrating strong detection performance across varying corruption intensities and network sizes. While our results do not replace legal or economic analyses, they offer an effective complementary tool for auditors and policymakers to monitor procurement integrity more effectively. This study demonstrates that simple geometric insight can reveal hidden dynamics in real-world networks better than other Information Theoretic metrics.

\end{abstract}

\begin{highlights}
   \item Heron’s coefficient quantifies structural imbalance in dynamic networks.
   \item We applied the measure to eight years of Brazilian health bidding data (2013–2021), revealing similar abnormal patterns before, during and after COVID-era procurement, possible indicators of collusion.
   \item Artificial network tests confirm anomaly detection stability when scaled.
   \item HIC supports auditing by highlighting unusual network motifs and links with greater sensitivity than other metrics.
\end{highlights}

\begin{keyword}
Complex Networks \sep Information Theory \sep Community Structure \sep Bidding and Tendering \sep Network Science \sep Covert Networks



\end{keyword}

\end{frontmatter}




\section{Introduction}

The outbreak of the COVID-19 pandemic in 2020 brought forth an unprecedented demand for medical supplies worldwide, causing potential risks to the availability of essential healthcare items \cite{teremetskyi_corruption_2021, steingruber_corruption_nodate, noauthor_exploitative_2020}. This surge in demand resulted in fierce competition among suppliers, particularly for individual protection materials such as hospital aprons, rubber gloves, and masks. In Brazil, where health is constitutionally guaranteed as a universal right, the government is responsible for the procurement and distribution medical supplies for over 210 million citizens through the Brazilian Unified Health System (SUS) \cite{paim_brazilian_2011, puppim_de_oliveira_brazilian_2017, gragnolati_twenty_2013, viana_universalizing_2017, collins_decentralising_2000,  almeida_health_2000, de_almeida_botega_brazilian_2020}.

Public procurement plays a crucial role in ensuring fair competition among companies to meet the nation's healthcare demands. However, unethical practices, such as collusion, have been observed among co-bidding companies, raising concerns about the integrity of the procurement process. Collusion in bidding networks is challenging to detect due to the complexity of such networks \cite{wachs_network_2019}.

To address these issues, we propose a measure called Heron's Information Coefficient (HIC) to gain insights into bidding processes for health-related items in Brazil. By understanding network structures and dynamics, we aim to identify potential collusion and enhance the transparency and integrity of public procurement. As such, we consider both the topological connectivity and the degree of protection provided to the system's elements. Complex network theory provides effective tools for analyzing various dynamic systems, enabling a better understanding of transportation, diffusion, synchronization, and social networks, among others \cite{albert_error_2000, aytac_network_2018, bachmann_survey_2020, otsuka_robustness_2019, smith_competitive_2019, tejedor_network_2017, wang_entropy_2006, wu_enhancing_2017, zhouInfluenceInterlinkTopology2020}.

Our approach focuses on network robustness, quantifying a network's resilience under failures or changes. By applying Heron's coefficient, we assess the system's protection level and its efficiency, considering the activity patterns of individual nodes or links. This approach can be extended to devise intelligent attack strategies and evaluate the impact of intentional failures on interconnected systems.

The proposed methodology was tested on both real and artificial networks, revealing insightful behavior in artificial networks and highlighting unusual patterns in the interactions among co-bidding companies for medical supplies in Brazil. The results of this study can provide valuable information to control agencies, aiding their supervision of public procurement processes and combating corruption effectively.

\section{Methods}

A network is a set of vertices, $V$, connected by links, $E$. Generally, the network is just the structure over which a time-dependent dynamical process occurs, and its elements must not be available to operate during all periods. In this way, at a given instant $t\geq0$, the original network $G=(V,E)$ can be decomposed into two components that represent the active and inactive subgraphs at that instant: $G^a(t)=(V^a(t),E^a(t))$ and $G^i(t)=(V^i(t),E^i(t))$.

\begin{figure}[h]
 \centering
 \includegraphics[width=0.7\linewidth,keepaspectratio=true]{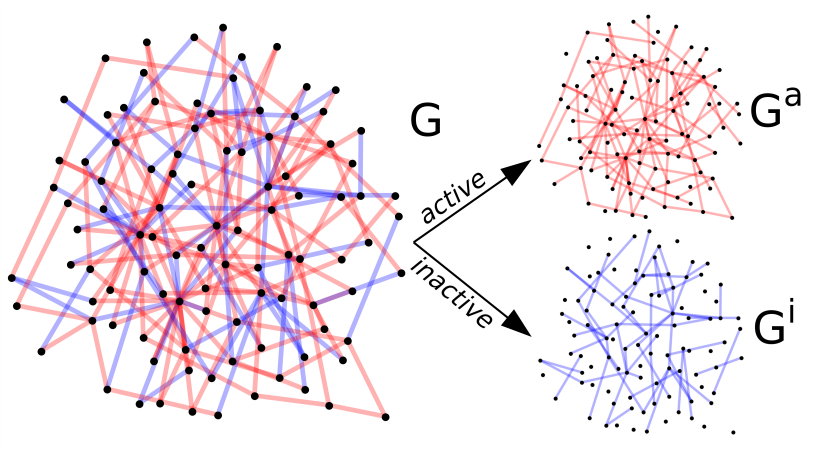}
 \caption{A small network $G$ with two complementary subgraphs $G^a$ (active) and $G^i$ (inactive). The D-distance between each pair is $D(G,G^i)=0.4582513$ $D(G,G^a)=0.3032006$ and $D(G^i,G^a)=0.2715595$. The largest triangle with perimeter $D(G,G^i)+D(G,G^a)+D(G^i,G^a)$ has area $0.05134147$ but the triangle with sides $D(G,G^i)$, $D(G,G^a)$ and $D(G^i,G^a)$ has $0.03964952$ and, thus, the fraction between the equilateral triangle and this second triangle $0.03964952/0.05134147=0.77$ is the Heron's coefficient.}
 \label{fig:active-inactive}
\end{figure}

If $G^a=G$, the entire network is active but very vulnerable to failures as any activity can fail. Conversely, the condition $G^i=G$ implies a system in topological stasis, fully protected yet dynamically inert. All other intermediate possibilities result in networks that may be more or less similar to the original one. To compare the active and inactive parts and the original network, we employ the $D$ measure \cite{schieber_quantification_2017}, a pseudo-metric between networks. This highly precise measure of network dissimilarity relies on the Jensen-Shannon divergence \cite{lin1991} to assess topological differences in connectivity patterns, effectively capturing the divergence in information flow along shortest paths.

Formally, let $P_i=\{p_i(j)\}$ denote the distance distribution of node $i$, where $p_i(j)$ represents the fraction of nodes located at a shortest-path distance $j$. We quantify the topological heterogeneity of the system via the Network Node Dispersion (NND), defined as $\mathrm{NND}(G)=\mathcal{J}(P_1,\ldots,P_N)/\log(d+1)$, where $\mathcal{J}$ is the Jensen-Shannon divergence and $d$ is the network diameter. The $D$-measure synthesizes three distinct topological descriptors: (i) the global difference between the averaged node-distance distributions, $\mu_G$ and $\mu_{G'}$; (ii) the aforementioned NND, which captures local variance in distance profiles; and (iii) a spectral comparison of the $\alpha$-centrality values of the graphs and their complements. This final term, also computed via Jensen-Shannon divergence, is critical for resolving isomorphisms in distance-regular structures and distinguishing between complete graphs of varying sizes or systems containing disconnected components. As such, the distance $D(G,G_0)$ between two networks enables a comparison based on global distance structure rather than local connectivity alone: 
\begin{equation}
\begin{multlined}
D(G,G') =
w_1 \sqrt{\frac{\mathcal{J}(\mu_G,\mu_{G'})}{\log 2}}
+ w_2 \big|\mathrm{NND}(G)-\mathrm{NND}(G')\big|
+ w_3 \left(
\sqrt{\frac{\mathcal{J}(P_G^{\alpha},P_{G'}^{\alpha})}{\log 2}}
+
\sqrt{\frac{\mathcal{J}(P_{G^c}^{\alpha},P_{G'^c}^{\alpha})}{\log 2}}
\right),
\end{multlined}
\label{eq:DGG}
\end{equation}

where $N$ and $M$ are the sizes of $G$ and $G'$, respectively, and $G^c$ indicates the complement of $G$. The weights $w_1$, $w_2$ and $w_3$ satisfy $w_1+w_2+w_3=1$, and the authors of the D-measure \cite{schieber_quantification_2017} show that the choice of the weights does not change the metric character and present a detailed discussion on the weight selection. Typically, $w_3$ is set to 0.1, as a high value for this term would distort the D values, minimizing the power of the NND in detecting these important topological changes. Due to the computational complexity added with this term, $w_3$ can can be avoided without losing precision - unlike $w_1$ and $w_2$, which are considered equally important and thus set to $w_1 = w_2 = 0.5$. By setting $w_3 = 0$, we rely entirely on terms $w_1$ and $w_2$, which are much better at detecting massive structural breaks, such as phase transitions or disconnections. 

In addition to its precision, the $D$-measure also satisfies the triangular inequality, so that, for any given three networks $G_1$, $G_2$ and $G_3$, $D(G_1,G_3)\leq D(G_1,G_2)+D(G_2,G_3)$. By satisfying positive definiteness, symmetry, and the triangular inequality, the $D$-measure effectively projects the complex, non-Euclidean topology of graph structures into a searchable metric space \cite{Klette_Rosenfeld_2004}. This projection ensures that the familiar laws of geometry and algebra remain valid, allowing us to reason about the distance between complex systems as rigorously as we reason about distance in the physical world.

The operational impact of this geometric validity cannot be overstated. The triangular inequality is the cornerstone of efficient high-dimensional search, since it enables both the computational scalability and the semantic definition of structural anomalies. It allows us to bound the distance between two objects without explicitly comparing them: if $G_1$ is close to $G_2$, and $G_2$ is far from $G_3$, the inequality mathematically guarantees that $G_1$ is also far from $G_3$. Without the pruning power granted by this axiom, the index structures required to identify outliers in billion-node graphs would be computationally intractable. Thus, the inequality transforms the concept of an "anomaly" from a subjective observation into a geometric fact: an object that physically cannot be embedded within the dense manifold of the distribution without violating the axioms of the space.

Given the triangular inequality, if at least one of these distances is positive, there are infinite different triangles possessing a perimeter ($2P$) equal to three times the average value between their distances: $2P=D(G_1,G_2)+D(G_2,G_3)+D(G_1,G_3)$. The most symmetrical situation occurs when the distance values are the same, resulting in an equilateral triangle in the metric space. By extension, all other configurations generate more heterogeneous distance patterns and the perimeter of their triangles becomes a measure of the total structural divergence within the triplet. Since the largest triangle for a given perimeter is equilateral and the Heron's formula gives the area of a triangle with sides $D(G_1,G_2)$, $D(G_1,G_3)$ and $D(G_2,G_3)$ as $\sqrt{P(P-D(G_1,G_2))(P-D(G_1,G_3))(P-D(G_2,G_3))}$, we define Heron's Information Coefficient as the fraction between the area of that triangle and the area of the equilateral triangle with perimeter $2P$:

\begin{equation}
\mathcal{H}_e(G_1, G_2, G_3) =
\frac{3\sqrt{3P \bigl(P - D(G_1, G_2)\bigr)
                \bigl(P - D(G_1, G_3)\bigr)
                \bigl(P - D(G_2, G_3)\bigr)}}{P^2}.
\end{equation}

\begin{figure}[h]
 \centering
 \includegraphics[width=0.9\linewidth]{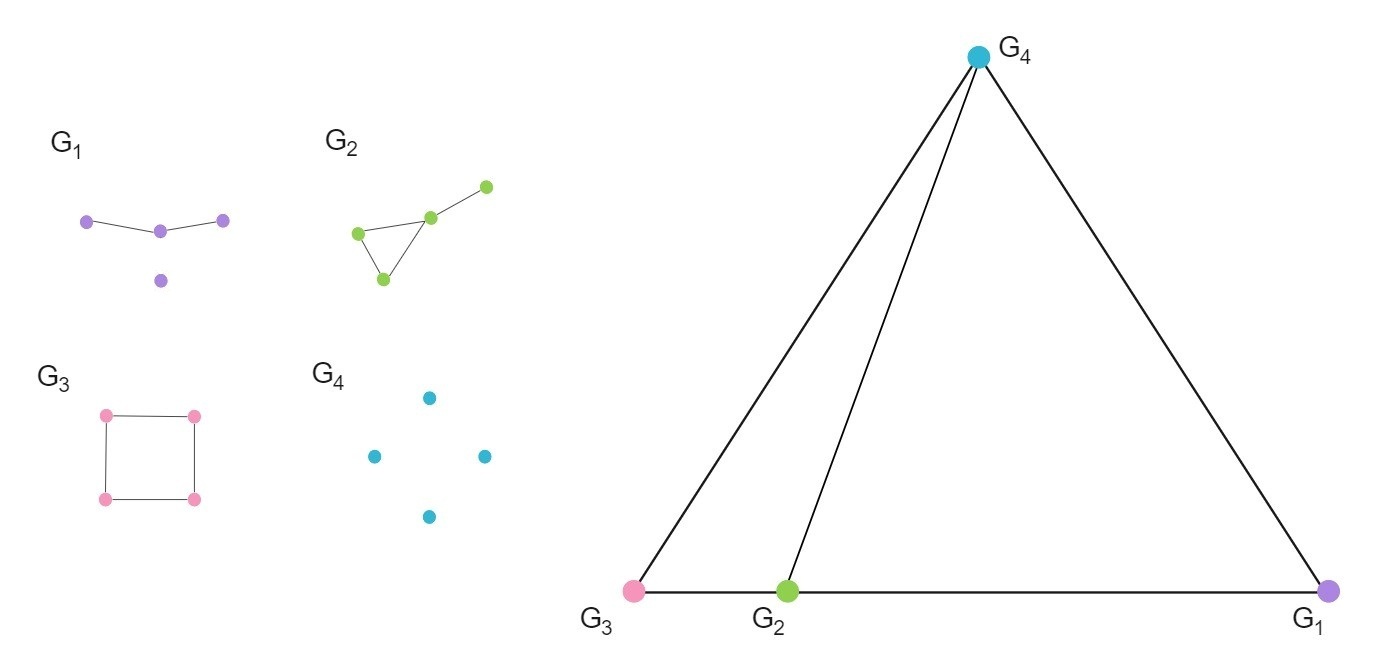}
 \caption{Multi-dimensional scaling map considering the $D$-distance and four small networks, $G_1$, $G_2$, $G_3$ and $G_4$. Here we visualize how non-isomorphic networks with 4 nodes form triangles with different areas based on the $D$-measure. }
 \label{fig:mds}
\end{figure}

Figure \ref{fig:mds} shows a multidimensional scaling map of the distances $G_1$, $G_2$, $G_3$ and $G_4$ between four small networks. In this representation, we can see that the largest possible triangle formed by triples of these networks has vertices $G_1$, $G_3$ and $G_4$ with similar sides. When combining $G_1$, $G_3$ and $G_4$ no triangle is formed because, in fact, $D(G_1,G_3)=D(G_1,G_2)+D(G_2,G_3)$. Thus, among all non-isomorphic networks of size four, ${\cal H}_e(G_1,G_3,G_4)=0.996$ represents its highest possible value while ${\cal H}_e(G_1,G_2,G_3)=0$ represents its minimum.

When HIC equals 1, the three networks' topologies in the system are equally different. In contrast, a zero value means a degenerate triangle, that is, either at least two topologies are equal or the topological differences between a pair are as large as the sum of the other two's topological differences.

\section{Characterizing Heron's Information Coefficient's Behavior}

\begin{figure}[h!]
 \centering
 \includegraphics[width=\linewidth]{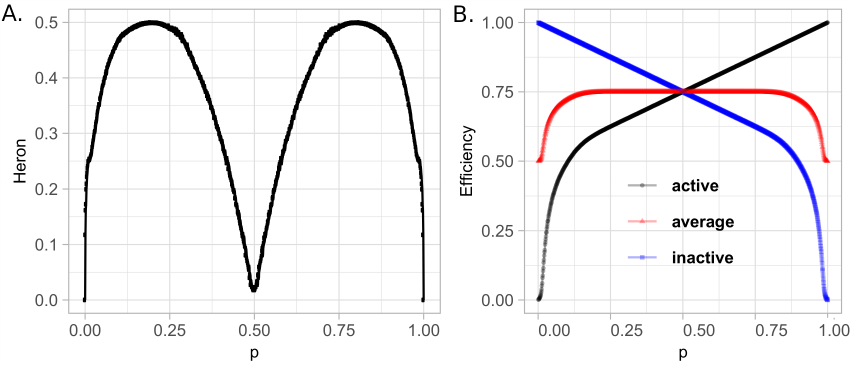}
 \caption{(A) shows the average values of the Heron's Information Coefficient for 100 experiments. (B) shows the network efficiency evolution for active and inactive subgraphs, as well as the average. The average efficiency of the active and inactive subgraphs tend to plateau close to 0.75 after a spike of the maximum HIC value.}
 \label{fig:er}
\end{figure}

To understand the role that Heron's Information Coefficient has as a measure of activity and protection of a networked system, we first consider a random graph model, Erdős-Rényi \cite{ErdosRenyi1960}. In this model, each link $i$ is independently activated according to a uniform distribution $p_i\in[0,1]$, where $p_i=0$ represents that link $i$ has no activity in the network, and $p_i=1$, its complete activity. Figure \ref{fig:er}(A) shows the average values of HIC for 100 experiments with different seeds. 

For small $p$ values, the inactive network closely resembles the complete graph, resulting in low HIC values. As $p$ increases, the topologies of the original, active, and inactive networks diverge until a maximum, reached when approximately 20\% of the links are active. Beyond this point, the HIC gradually decreases until a local minimum in $p=0.5$. At this local minimum, the active and inactive networks are topologically similar subgraphs of the original network, as the probability of the existence of a link is 50\% in both.

The random graph model is a good example, but it does not tell the whole story. In such a network, each link has an equal probability of being active, a pattern that is not often presented in the real world. In social networks, for example, some ties tend to be stronger than others. To introduce such heterogeneity while maintaining the same expected number of links, we assign link activation probabilities proportional to their edge betweenness centrality ($b_e$), a measure reflecting the number of shortest paths that traverse each edge. Specifically, for a given parameter $\gamma$, the activation probability of an edge is defined as $p_e \propto  b_e^\gamma$. 

\begin{figure}[h]
 \centering
 \includegraphics[width=\linewidth]{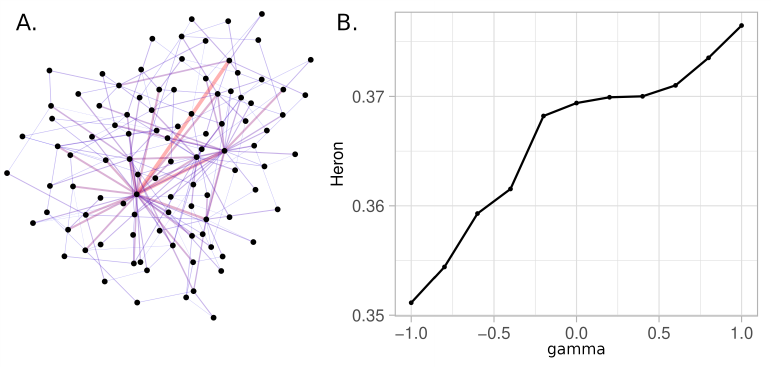}
 \caption{Network with 10\% of active links, where link activation probability depends on edge betweenness centrality ($b_e$) and a real number $\gamma$, such that $p_e \propto  b_e^\gamma$ being proportional to $b_e^\gamma$. (A) shows an example of active and inactive edges and (B) the average HIC for 100 seeds across different $\gamma$ values.}
 \label{fig:atin}
\end{figure}

Figure~\ref{fig:atin}(A) shows an example of an active–inactive configuration in a network with 100 vertices, where the thicker and redder edges denote higher betweenness centrality values. As depicted in Figure~\ref{fig:atin}(B), we computed the average HIC across 100 random seeds for different $\gamma$ values. Negative  $\gamma$ values correspond to preferential activation of low-betweenness edges, whereas positive  $\gamma$ values emphasize highly central ones. The results reveal that HIC tends to increase when edges with higher betweenness are more likely to be active, reflecting a more efficient structural differentiation between active and inactive subgraphs.

Overall, a high Heron coefficient is associated with the system's efficiency in transferring information. We observe high values in systems whose active and inactive parts are equidistant from the original network and from each other, which indicates different connectivity patterns without losing information from the original network. In the network represented in Figure \ref{fig:atin}, the preferential attachment to high edge betweenness edges ensures that the information transfer during the activity period travels the network more efficiently in different points, whereas inactivity patterns protect the network's periphery. Simultaneously, however, important links in the system - those with high betweenness centrality, for instance - may fail. In this case, a possible strategy would be to interchange the active and inactive networks while retaining Heron's information coefficient so that the information that reaches the system's periphery rapidly through the most central active network can further spread to peripheral nodes. 

This analysis highlights that edges connecting distinct network modules, which are typically those with high betweenness, play a crucial role in maintaining global connectivity. Prioritizing their activity preserves inter-community communication and enhances information flow across the network. Notably, betweenness centrality is among the most effective measures for identifying strategic positions within covert or collusive structures \cite{cavallaroDisruptingResilientCriminal2020, ficaraCovertNetworkConstruction2022}. Furthermore, this centrality is used to detect collusion through sequential edge removal, in a process analogous to a targeted network attack. 

As such, we next evaluate HIC’s sensitivity to network perturbations relative to classical robustness metrics: degree centrality, betweenness centrality, clustering coefficient, and modularity. The goal is to assess how effectively each metric captures a network’s ability to preserve connectivity and structural integrity under deliberate disruptions.

To this end, we conducted experiments across multiple network topologies. The scale-free Barabási–Albert model grows through preferential attachment, producing hubs with high degree centrality and a topology resilient to random failures but highly vulnerable to targeted attacks on hubs. The small-world or Watts–Strogatz model combines local clustering with short global path lengths, which confer it balanced robustness across attack strategies. On the other hand, the Erdős-Rényi network, whose robustness depends mainly on edge density, served as a baseline due to its rapid fragmentation under targeted attacks. We also analyzed a community-structured network, characterized by densely connected clusters with sparse inter-community links, to evaluate the effects of bridge removal on global connectivity. In contrast, the hub-and-spoke topology is a centralized structure dominated by a single hub, extremely fragile under targeted removal of that central node, emphasizing dependence on a single point of failure. Finally, the real bidding network exemplifies the problem we aim to tackle, capturing the genuine structural patterns observed in procurement systems. 

\begin{figure}[h!]
 \centering
 \includegraphics[width=\linewidth,keepaspectratio=true]{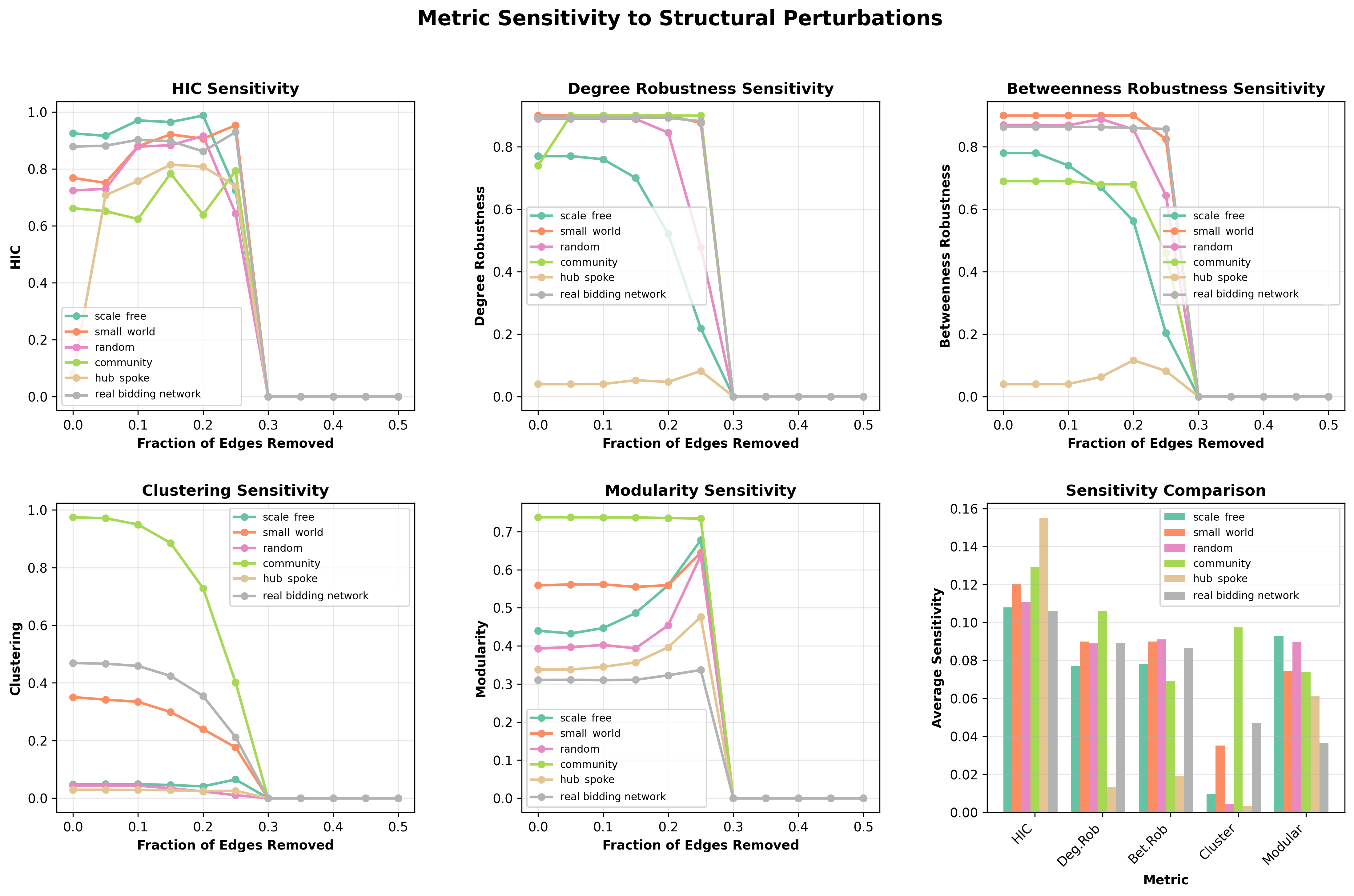}
 \caption{Sensitivity comparison of the edge removal across relevant metrics (HIC, degree robustness, betweeness robustness, clustering, and modularity) for different network types (scale-free, small-world, Erdős–Rényi, community-structured, hub-and-spoke, and a real bidding network). The real network corresponds to the 2016 Surgical Masks network, where our metric outperformed the others, as it consistently did for Brazilian bidding networks.}
 \label{fig:comp}
\end{figure}

Each network underwent controlled node removal experiments, in which 10\% of the nodes were sequentially removed in descending order of degree. Across all types of network, our coefficient consistently exhibited a higher sensitivity to structural degradation than the other metrics evaluated. Our coefficient's performance reinforces its suitability for detecting covert or collusive groups, surpassing state-of-the-art methods based on edge betweenness in capturing critical structural vulnerabilities.

Extensive testing across multiple network configurations confirmed the expected behavior and robustness of our coefficient. Together, these findings demonstrate that Heron’s Information Coefficient captures both the structural balance between active and inactive subgraphs and the dynamic efficiency of the information flow within the system. Its sensitivity to perturbations across a variety of topologies highlights our coefficient's potential.

\subsection{Heron's Information Coefficient in collusion detection}

Within the broader anomaly detection literature, detecting anomalies in data serves as a critical filter for identifying deviations in high-dimensional systems, spanning domains from financial forensics to social network security. While applicable to various data modalities, we prioritize graph-based approaches due to their unique capacity to model long-range correlations and structural interdependencies. In the context of fraud, for instance, anomalies are rarely isolated events; they are inherently relational. For instance, fraudulent activity typically manifests through specific topological patterns: either as opportunistic contagion, where fraud propagates through established social links; or as organized collusion, where dense subgraphs of actors synchronize their behavior \cite{akogluGraphBasedAnomaly2015}. 

Historically, detection frameworks relied on heuristic feature engineering and domain-specific statistical models. This paradigm is labor-intensive but also brittle, struggling to generalize to previously unseen anomaly types \cite{maComprehensiveSurveyGraph2023}. While traditional machine learning techniques, such as Support Vector Machines (SVMs) and matrix factorization, attempted to automate feature extraction, they frequently succumb to the curse of dimensionality and prohibitive computational costs when scaling to massive adjacency matrices. Consequently, the field has pivoted toward Deep Learning to capture the non-linear complexity of graph data and learn expressive latent representations capable of separating normal from anomalous manifolds. However, this gain in predictive performance introduces a new critical limitation: these models operate as "black boxes," offering superior accuracy but lacking the interpretability required for high-stakes decision-making.

More specifically, the study of covert networks is challenging due to a myriad of reasons, one of which is data incompleteness and unreliability. Publicly available and reliable datasets, such as the Sicilian Mafia dataset \cite{cavallaroDisruptingResilientCriminal2020}, are rare and heavily contingent on judicial disclosures. Criminal network research is also uniquely sensitive to sampling biases. For example, random sampling can severely distort the understanding of a network clustered around a few focal nodes. In fact, the collection of complete data is a virtually impossible task, failing to properly capture the complete dynamics of the networks, with bias being inevitable \cite{ficaraCovertNetworkConstruction2022}. This uncertainty is compounded by the boundary-specification problem: in the absence of explicit membership rosters, defining the limits of a clandestine network requires inferring latent connections within a secretive, non-cooperative environment.

Moreover, the current detection methods used by law enforcement agencies are restricted from the greater public. Understandably so. Knowing the exact criteria used to identify potential criminal agents could ease organized crime's efforts in evading capture, in an adversarial adaptation. In the criminology literature, a recurrent approach consists of Network Science techniques. Its primary utility lies in its ability to map dynamic covert systems without relying on \textit{a priori} organizational templates. Rather than forcing data into presumed hierarchies, graph-theoretical approaches allow the structural properties of the network to emerge from the empirical data.

However, the interpretation of these emergent topologies requires significant nuance to avoid misleading inferences. For instance, the high degree connectivity of an element in the network does not necessarily equate to high status and, conversely, peripheral actors, with low connectivity, may exercise disproportionate control through strategic compartmentalization. For instance, in human trafficking rings, "madams" may appear peripheral because they operate independently, despite driving the market demand and controlling "managers" who appear more central in terms of degree centrality \cite{campanaExplainingCriminalNetworks2016}. Therefore, structural measures must be contextually grounded. Researchers must must integrate edge attributes and qualitative metadata, recognizing that the absence of a tie carries as much information-theoretic weight as its presence.

Notably, sequential node removal based on betweenness centrality is among the most effective measures for identifying strategic positions within covert or collusive structures \cite{cavallaroDisruptingResilientCriminal2020, ficaraCovertNetworkConstruction2022}. Betweeness produced a greater impact in dismantling a covert network, showing its capability to properly identify relevant elements. This evidence underscores that in covert networks, the capacity to bridge structural holes and control information flow is a more accurate predictor of systemic importance than mere connectivity.

For bidding processes, the collusion between companies is hard to identify and can take multiple forms, using the techniques mentioned in this section \cite{garciarodriguezCollusionDetectionPublic2022}. For instance, the recurrence of bids for a particular item {\cite{steingruber_corruption_nodate, vian_review_2008}} can contribute to cartel companies' prior planning and their subsequent organization in rounds and punishment of the players who fail to cooperate in the next bid. Companies that act in collusion usually participate in a large amount of bidding to share the average earnings or confer an impression of honesty throughout the process {\cite{vian_review_2008}}. Therefore, we model each bidding item as a network in which each node is a company participating in the bidding process. Each link exists if two companies are competitors in the same contest. To that end, multiple links represent multiple instances where the companies were co-bidding. As such, the bidding process possesses a complex topology structure that is very informative but often difficult to analyze. 

The theoretical benchmark of a public bidding network consists of nodes uniformly connected and characterized by the nonexistence of strong and connected communities, leading to a challenging trade-off: extracting connections from this set and retaining those that represent a high-level interaction between companies without overemphasizing what could be simple statistical deviations. In this scenario, Heron's Information Coefficient is useful due to its quantification of structural imbalance in dynamic networks, revealing signs of the coordinated behavior typical of collusion. 

\begin{figure}[h!]
 \centering
 \includegraphics[width=\linewidth,keepaspectratio=true]{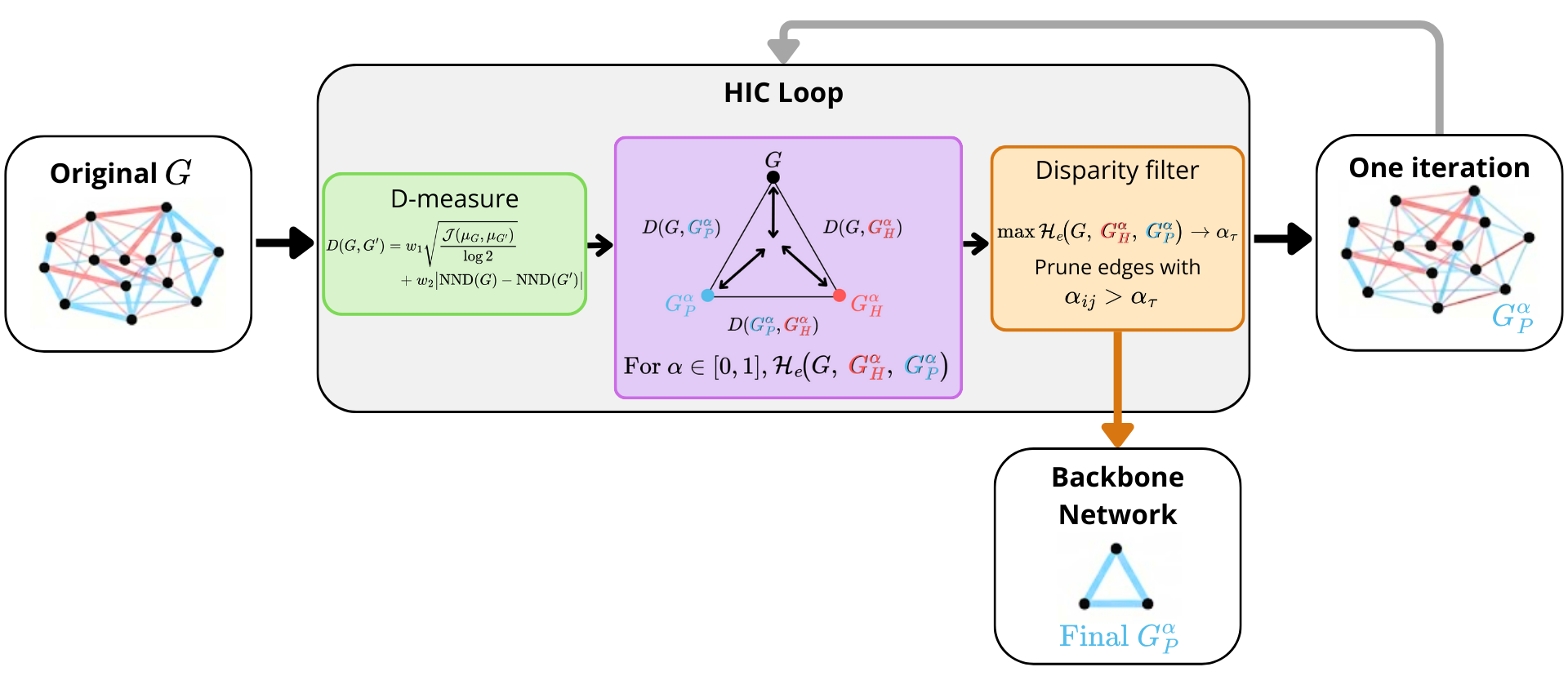}
 \caption{Diagram of our HIC method for collusion detection with a disparity filter. The HIC loop consists of finding the optimal separation threshold $\alpha_T$, defined as the point where the structural divergence between the original network ($G$), the backbone ($G_P^\alpha$), and its complement ($G_H^\alpha$) is maximized. Geometrically, this corresponds to maximizing the area of the triangle formed by these three states in the D-measure metric space in each iteration. HIC is maximized for a set number of iterations or until the graph is entirely disconnected, depending on the experiment.}
 \label{fig:diagram}
\end{figure}

Here, we use the disparity filter algorithm \cite{serranoExtractingMultiscaleBackbone2009} alongside the maximization of HIC to highlight the central players in the BID network. Figure \ref{fig:diagram} shows how our method works iteratively.

The idea of the disparity filter is to associate with each network's link a weight $\alpha_{ij}$ that represents a significant level provided by the distribution of the local uniform randomness null hypothesis \cite{serranoExtractingMultiscaleBackbone2009}. Relevant links are chosen so that $\alpha_{ij}<\alpha$, $\alpha \in[0,1]$ being a fixed significance level. Instead of arbitrarily choosing the significance levels, we maximize HIC when $\alpha$ varies. Disconnected companies are excluded from the resulting network. The process is repeated until reaching a significance level as small as desired. The selection of the most relevant group is made iteratively, starting with the original network with a significance level equal to 1. A new subgraph is obtained at each step by removing edges that possess a higher significant level. The goal is to find a lower significance level for each step, resulting in a higher HIC. By definition, this happens when the topologies of the active, inactive, and original networks are close to equidistant. 

Each company that participated in the bidding process is a node connected by links weighted by the number of times they have competed in the same bidding process. The way nodes interact and connect provides essential information about the system {\cite{akbarpour_diffusion_2018}}. In particular, information about their connectivity patterns represents a valuable tool on how different dynamics occur in the system. For example, the speed at which information travels within a cohesive community is much greater than in a more sparse one. As such, let $G$ be the original weighted network after applying the disparity filter, for a fixed ${\alpha}\in[0,1]$, and $G_a^{\alpha}$ be the network obtained from $G$ by removing links whose weight is greater than or equal to $\alpha$. As such, $G_i^\alpha=G-G_a^\alpha$ is the complement of $G_a^\alpha$ to $G$. We define $\alpha_T$ as the smallest value of $\alpha$ such that ${\cal H}_e(G,G_a^\alpha,G_i^\alpha)$ is as large as possible. This optimal threshold $\alpha_T$ thus identifies the most informative backbone of the network, balancing cohesion and separation between active and inactive subgraphs. In doing so, it reveals the most representative organization of competitive interactions among companies.

To characterize HIC's behavior in real procurement data, we use a Brazilian dataset of infrastructure projects procurement \cite{garciarodriguezCollusionDetectionPublic2022}, which documents the procurement activities of the state-controlled oil company Petrobrás between 2002 and 2013. This era encompasses the systemic market manipulation uncovered by the "Operation Car Wash" investigation, a landmark inquiry that exposed the "Club of 16" - a cartel of Brazil's largest construction firms that orchestrated bid-rigging and price - fixing across major infrastructure projects. Despite lacking the temporal aspect, this dataset provides valuable insights from known collusive agents in Brazilian biddings. The original authors established, between machine leaning methods, the superior efficacy of ensemble methods, specifically identifying Extra Trees, Random Forest, and AdaBoost as the top-performing algorithms, with detection rates between 81\% and 95\%. We reproduced their experiments and tested disparity filters with HIC and degree robustness, given its strong performance in the experiments shown in Figure \ref{fig:comp}.

\begin{figure}
    \centering
    \includegraphics[width=0.8\linewidth]{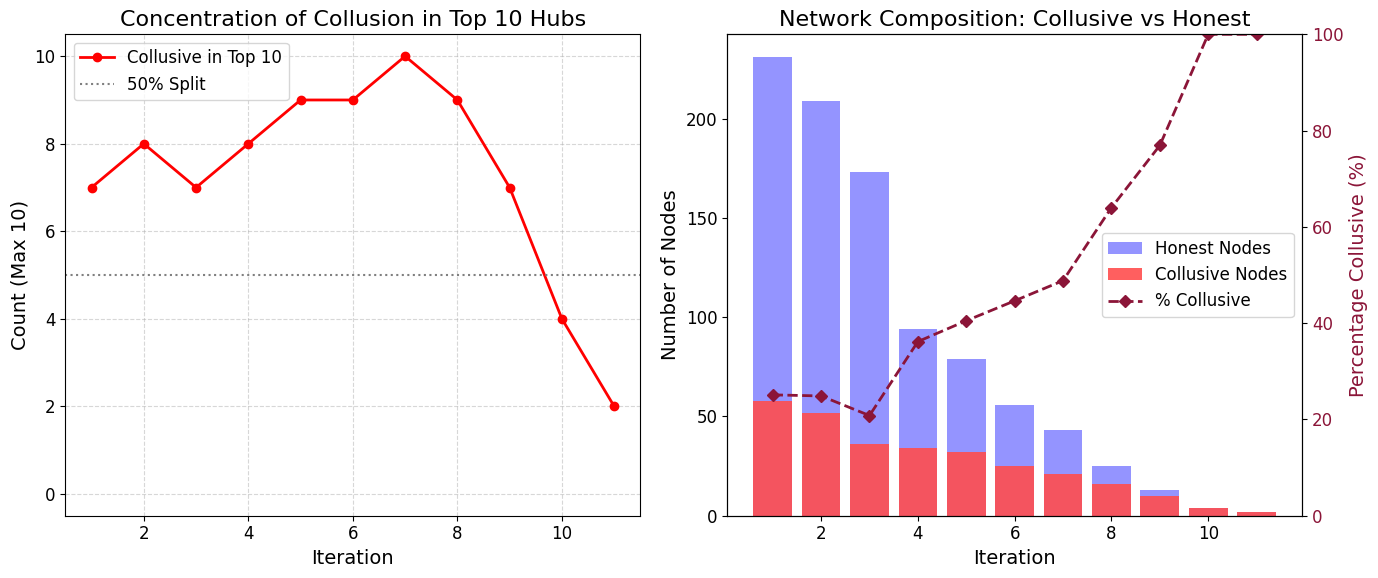}
    \caption{Evolution of network composition during iterative backbone extraction for Petrobrás procurement from 2002 to 2013. (Left) Hub Dominance: The count of collusive agents within the top 10 most central nodes identified with HIC, across iterations. The algorithm consistently retains collusive "ringleaders" in the top ranking, peaking at 100\% dominance by Iteration 7. (Right) Network Composition: The stacked bars show the absolute number of nodes remaining at each step, colored by ground truth (Blue = Honest, Red = Collusive). The dashed line tracks the percentage of collusive nodes in the remaining subgraph. As the algorithm iterates, it disproportionately sheds honest nodes (noise), resulting in a monotonic increase in cartel density, reaching 100\% purity at Iteration 10.}
    \label{ent}
\end{figure}

The network comprises 272 corporate nodes, of which 47 (17.28\%) are identified as corrupt; notably, only 80 companies successfully won at least one of the 101 analyzed auctions. Within the total volume of 683 bids, 128 were flagged as collusive. A defining structural characteristic of this network is its low collusion assortativity ($r = 0.1480$). This suggests a deliberate "cover bidding" strategy, where collusive agents dilute their presence among honest participants to mask topological clustering. Despite this challenging scenario, HIC was able to isolate only collusive agents after 10 iterations, as shown in Figure \ref{ent}. The only moment our method failed to identify at least 5 collusive agents in its top 10 was in the tenth iteration, when the network only had 4 companies - all of which were corrupt.

To validate these findings, we conducted a rigorous sensitivity analysis using Monte Carlo simulations with bootstrap sampling for stability and null models for significance testing. The results reveal a sharp divergence in metric utility. Standard topological measures, specifically Degree Robustness and Modularity, failed to reject the null hypothesis, suggesting they are blind to the cartel's non-modular structure. In contrast, HIC identified a significant structural anomaly (Z=4.32,$p<0.001$). The observed HIC value ($0.856$) lies far outside the null distribution mean ($\mu \approx 0.69$), confirming that the collusion manifests not as simple clustering, but as a subtle "information backbone." This provides compelling statistical evidence that HIC is a superior metric for detecting the subtle coordination characteristic of collusion, which standard topological measures overlook.

\section{Experimental Results}

All the results on this section are based on the extraction of the network's backbone with a disparity filter based on HIC maximization. The source code for the characterization of our coefficient's behavior and other experiments can be found on GitHub (\href{https://github.com/FutureLab-DCC/Heron_coefficient}{https://github.com/FutureLab-DCC/Heron\_coefficient}).

\subsection{Artificial bidding network}

We tested our method in artificial procurement networks generated using the Bernoulli Weighted Random Network (BWRN) model \cite{elsisyNetworkGeneratorCovert2022}, which combines the Stochastic Block Model (SBM) for group formation with hierarchical structures for management layers. This generator creates synthetic networks that are statistically equivalent substitutes for real networks, allowing for research on covert networks. Its validity has been established through comparison with the publicly available Sicilian Mafia network \cite{cavallaroDisruptingResilientCriminal2020}, which is a covert network that covers phone calls and in-person meetings extracted from court proceedings. Thus, these synthetic networks replicate the behaviors characteristic of real colluding agents while addressing a critical challenge in covert network research: data availability \cite{campanaExplainingCriminalNetworks2016}. We enhanced realism by incorporating successful bids to better simulate procurement networks with known corrupt agents. Then, we iterated the HIC filtering process until network sparsity prevented further analysis.

\begin{figure}
    \centering
    \includegraphics[width=\linewidth]{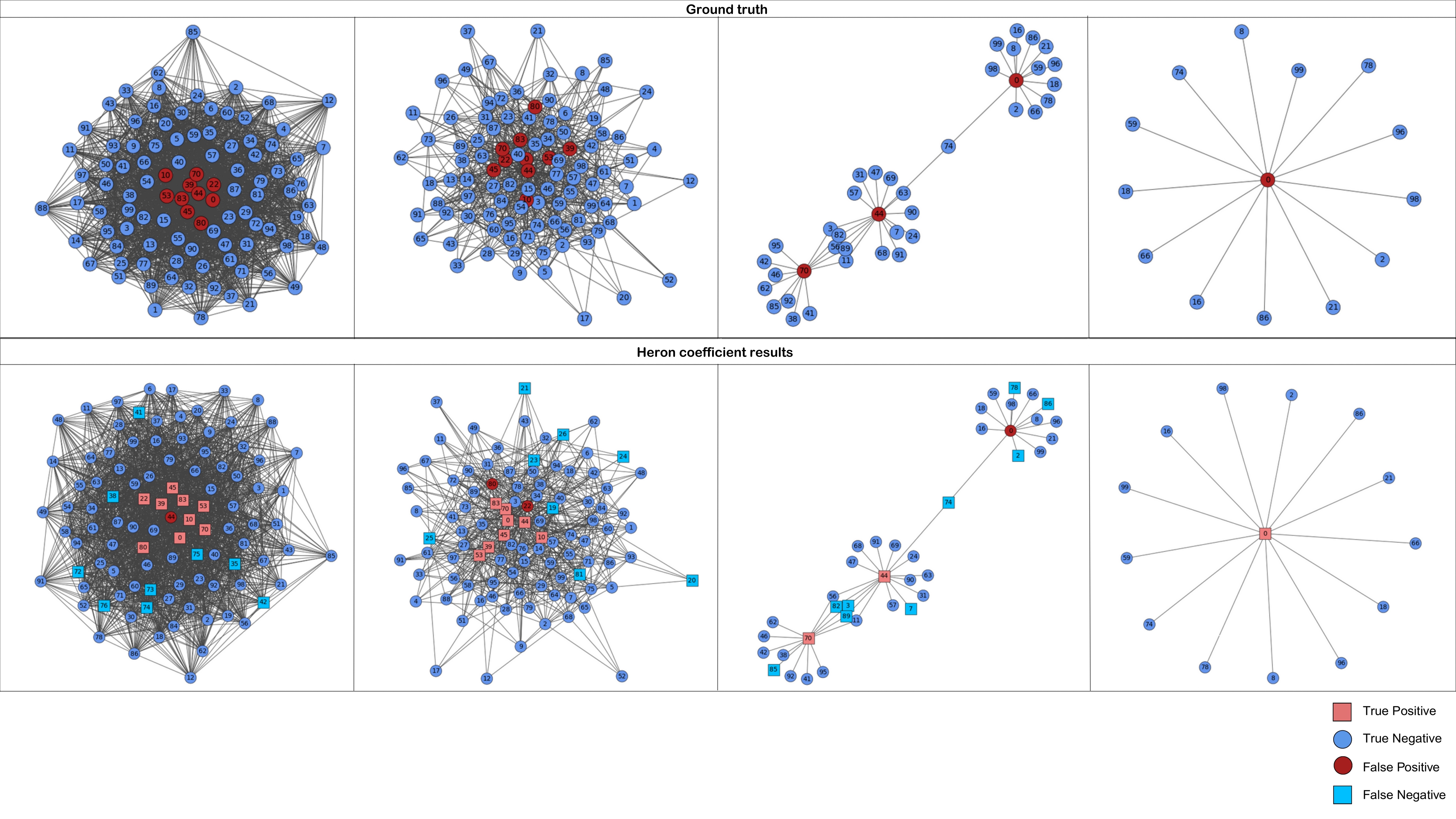}
    \caption{Disparity filter with HIC to detect corrupt agents in a synthetic network. The image shows iterations 1, 3, 5 and 6 (the last one) for 10 corrupt agents in a network of 100 agents. Above are the original networks with the correct labels, with the analyzed network of the iteration bellow.}
    \label{fig:networks}
\end{figure}

Figure \ref{fig:networks} demonstrates how the Heron coefficient progressively identifies corrupt agents. As the algorithm iterates, we can see that important corrupt nodes are maintained. These results corroborate our method's potential to detect colluding agents in real-world procurement data.

\begin{figure}[h!]
    \centering
    \includegraphics[width=\linewidth]{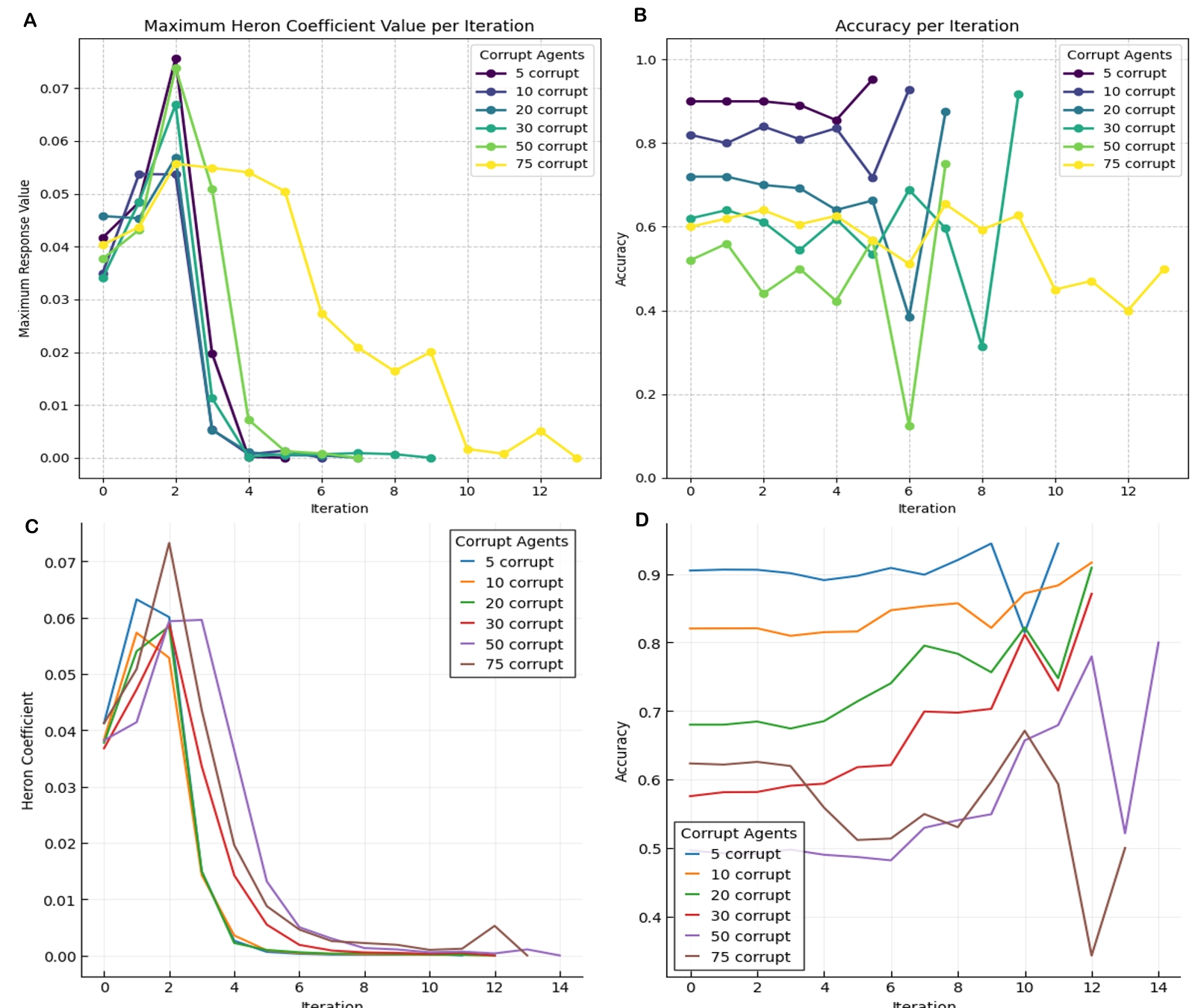}
    \caption{Comparison of different numbers of corrupt agents for networks with 100 agents, for 1 seed (A and B) and the average on 100 seeds (C and D) showcasing the method's stability. In later rounds, our method performs well even in challenging scenarios such as those where the deviating behavior is not isolated in a minority of agents, but in its majority.}
    \label{fig:acc}
\end{figure}

To validate robustness, we tested the method in 100 random seeds and varying parameters, iterating until all HIC values reached zero, that is, the network dissolved.  The results were consistent, as seen in \ref{fig:acc}. With a lower percentage of corrupt agents, the method stopped faster. Networks with lower corruption levels (5\%, 10\%) maintained higher accuracy throughout, generally staying above 0.8. Furthermore, the method showed great final accuracy for networks containing up to 30\% corrupt agents, with the final iteration successfully identifying at least one corrupt agent in most cases. Performance remains stable across network scales, although optimization would be necessary for graphs exceeding 10,000 nodes.

\begin{figure}[h!]
    \centering
    \includegraphics[width=0.6\linewidth]{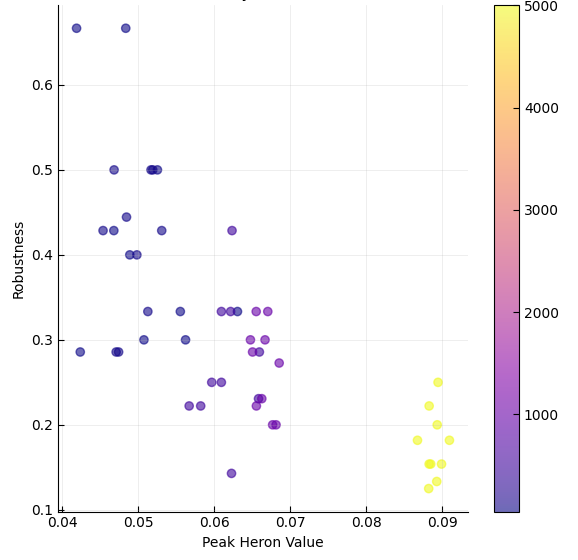}
    \caption{Comparison of robustness and peak HIC value for BWRN networks with 10\% of corrupt nodes, from 10 to 5000 corrupt agents. This scaling test demonstrates the effectiveness of using HIC to expose the critical backbone of covert networks.}
    \label{fig:scale}
\end{figure}

Scaling experiments on networks with up to 5,000 nodes (all generated with 10\% corrupt nodes across 100 seeds with the BWRN generator) revealed a fundamental structural principle. As shown in Figure \ref{fig:scale}, larger organizations exhibit reduced structural robustness during backbone extraction, precisely because our method successfully identifies their sparse, efficiency-optimized architectures. This inverse relationship demonstrates that covert networks create fragile organizational skeletons that rapidly fragment when key nodes are removed. The HIC's sensitivity to these optimized structures confirms its effectiveness in exposing the critical trade-off between scale and vulnerability in covert network design.

\begin{figure}[h!]
    \centering
    \includegraphics[width=0.6\linewidth]{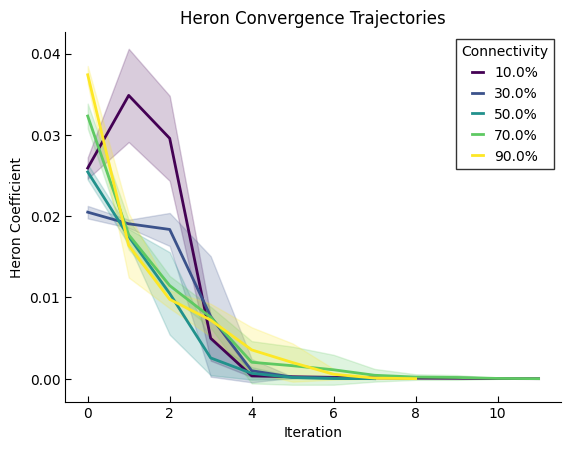}
    \caption{Average HIC convergence in incrementally growing networks from 10\% to 100\% of their final size, over 100 seeds. The networks were generated with the BWRN generator to be ultimately composed of 10 corrupt nodes among 100 total agents.}
    \label{fig:sliding}
\end{figure}

Our method also performs effectively with partial network information, as shown in Figure \ref{fig:sliding}. We tested on networks growing to have 10 corrupt nodes and 100 overall nodes in 10\% increments until their final size. We found that the average HIC behavior is consistent with that seen on Figure \ref{fig:er}. For 10 to 30\% of the network, we have a higher HIC and an early detection. On the other hand, we have a steep decline for 70 to 90\% of the network available. This trajectory suggests these networks possess a well-defined backbone where a subset of critical bidding relationships captures the system's essential dynamics. 

In summary, our method is computationally cheap, converges fast, and shows promising results in detecting corrupt agents in bidding processes. The Heron Information Coefficient effectively identifies corrupted agents in early iterations, particularly at lower corruption levels. While accuracy becomes less stable at corruption levels exceeding 50\% - an expected limitation since anomaly detection inherently struggles when anomalous nodes constitute the majority - the method remains valuable. Even after six iterations in high-corruption scenarios, it substantially reduces network size while retaining corrupt nodes. As such, our results suggest that the Heron coefficient can serve as a practical aid for corruption inspection, even in resource-constrained environments with limited detection capabilities.

\subsection{The BID Network}

Brazil is a country of continental dimensions, the fifth largest in the world and the sixth in terms of population \cite{noauthor_ibge_nodate} with considerable regional and social inequalities \cite{de_almeida_botega_brazilian_2020, almeida_health_2000, paim_brazilian_2011}. As determined by the country's Constitution, the provision of services in healthcare and education is a citizen's right, met with resources collected through taxation at the federal, state, and municipal levels \cite{paim_brazilian_2011, gragnolati_twenty_2013, de_almeida_botega_brazilian_2020, viana_universalizing_2017}. Supplied by public procurement, the national unified healthcare system (SUS) has promoted access to primary and emergency care and reached universal coverage of vaccination and prenatal care. As this system has become increasingly weak over the decades, the current infrastructure of primary care clinics and emergency units is mainly public. In contrast, hospitals, outpatient clinics, and diagnostic and therapeutic services are mostly private {\cite{puppim_de_oliveira_brazilian_2017, de_almeida_botega_brazilian_2020}}. Consequently, even before a strenuous situation on the health system, such as a pandemic, the decreasing public health expenditures are insufficient and underfunded {\cite{puppim_de_oliveira_brazilian_2017, paim_brazilian_2011}}. Furthermore, the inadequate funds allocated to the system are severely undermined by corruption in public procurement. The diversion of funds meant to address social needs contributes to the increase in poverty and inequality {\cite{teremetskyi_corruption_2021, vian_review_2008}} in an already unequal country{\cite{almeida_health_2000}}. Even with these setbacks, SUS has been able to attend to the population {\cite{gragnolati_twenty_2013, puppim_de_oliveira_brazilian_2017, paim_brazilian_2011}}, albeit with execution flaws often engendered by the ultimately political challenges the system faces {\cite{viana_universalizing_2017, mattos_procurement_2016}}. 

In 2020, the world faced a pandemic that urged the acquisition of essential and high-demand goods {\cite{noauthor_exploitative_2020, teremetskyi_corruption_2021}}, such as personal protection equipment. Consequently, the COVID-19 outbreak highlighted significant inconsistencies in national plans to contain a pandemic and the lack of good governance and corruption prevention {\cite{steingruber_corruption_nodate, teremetskyi_corruption_2021, kirya_anti-corruption_nodate}}. The response of the Brazilian system to the pandemic also faced political challenges {\cite{puppim_de_oliveira_brazilian_2017}} such as significant investments in the purchase of more than 5 million units of an ineffective treatment of the disease, chloroquine {\cite{wesseljun_22_its_2020, minas_exercito_2021, darlington_how_nodate}}. This context only exacerbates a historical problem in public procurement around the world: an estimated loss of 10\% to 25\% of global spending is due to corruption {\cite{kohler_does_2015, vian_review_2008}}.

Public procurement in healthcare is a significant area of risk of corruption \cite{vian_review_2008}, which can occur through bid-rigging, funds embezzlement, opacity in governance, misuse of power, nepotism and favoritism in management, petty corruption in levels of service, fraud and theft of medical resources {\cite{teremetskyi_corruption_2021}}. One such corrupt practice is cartel formation, which exposes the government to conspiring agents who violate the objective of bidding, incurring a loss of efficiency in public biddings on price and welfare in varying degrees {\cite{harrington_detecting_2005, vian_review_2008, teremetskyi_corruption_2021, steingruber_corruption_nodate}}. Although descriptive analyzes can provide information on how susceptible to fraud the Brazilian procurement system is, as shown by the high price variability for some bidding items \cite{harrington_cartel_2006}], we obtain little information on how the companies participating in a bidding process act together to bypass the inspection system. 

\begin{figure}[h!]
 \centering
 \includegraphics[width=.7\linewidth]{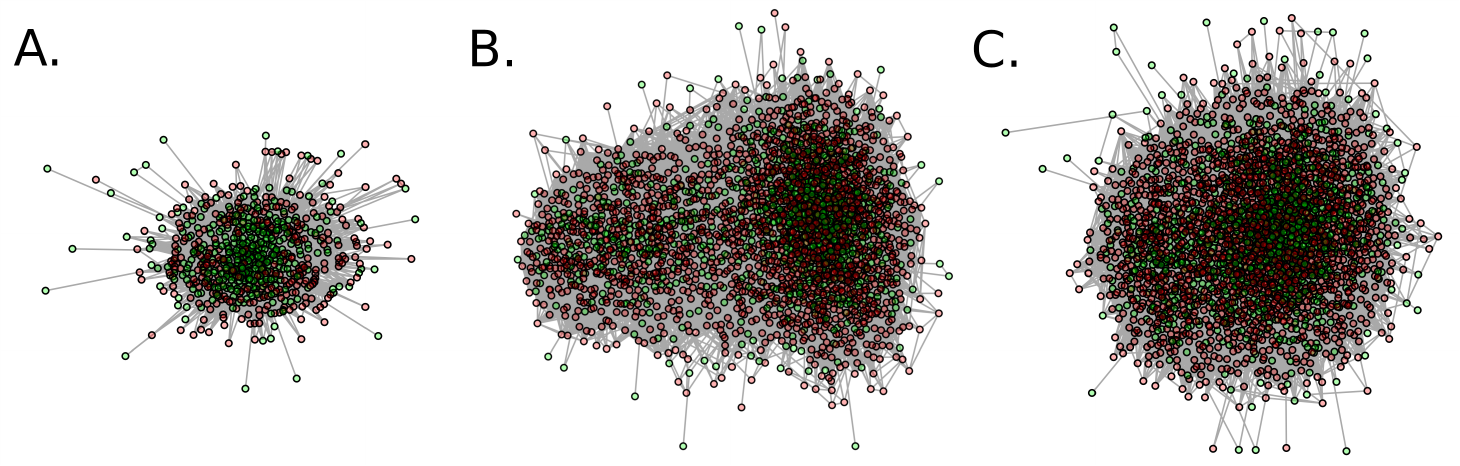}
 \caption{(A) Tube (B) Mask and (C) Glove between 2013 and 2019. In each graph, the green nodes are the companies that won at least one bidding process, whereas the red ones never won.}
 \label{fig:empresas}
\end{figure}

We consider data from the Brazilian Federal Government's public bids from 2013 to 2021 for three personal protective equipment: a surgical mask, rubber gloves and endotracheal tubes. The data is freely available through the Transparency Portal \cite{controladoria-geral_da_uniao_licitacoes_nodate}. This channel monthly publicizes information gathered from several Federal Government systems. Considering all the bids of a given item, the network is composed of nodes weighted by the number of participating bids, linked by edges weighted by the number of times two companies took part in a bidding process. We tested the networks of these three products over six steps. Figure \ref{fig:empresas} shows the networks for each one of the items, and a more detailed look at the companies year by year can be found in the Supplementary Material \ref{S1}.

The market for each item has some peculiarities. Masks were in great demand by institutions and individuals during the pandemic period, with a massive increase in demand and offer. Conversely, gloves are items with a relatively predictable annual order, used exclusively by healthcare workers. Despite the pandemic, canceling elective surgeries can correlate to maintaining this market's stable behavior.

It is important to note that a few companies hugely dominate the public bidding market for these items. Of the 733 companies participating in public bids for endotracheal tubes, 55\% won at least one process. However, a surprising 3.5\% won half of all submissions. Of the public proposals for masks and gloves, these numbers are 4\% and 2\%, respectively. Since the internal procedures of Brazilian bidding supervision agencies are, understandably, not publicly available, our approach relies on metrics and methods from covert network analysis, focusing on the sequential removal of edges \cite{cavallaroDisruptingResilientCriminal2020, ficaraCovertNetworkConstruction2022} via the disparity filter.

\begin{figure}[h!]
 \centering
 \includegraphics[width=\linewidth]{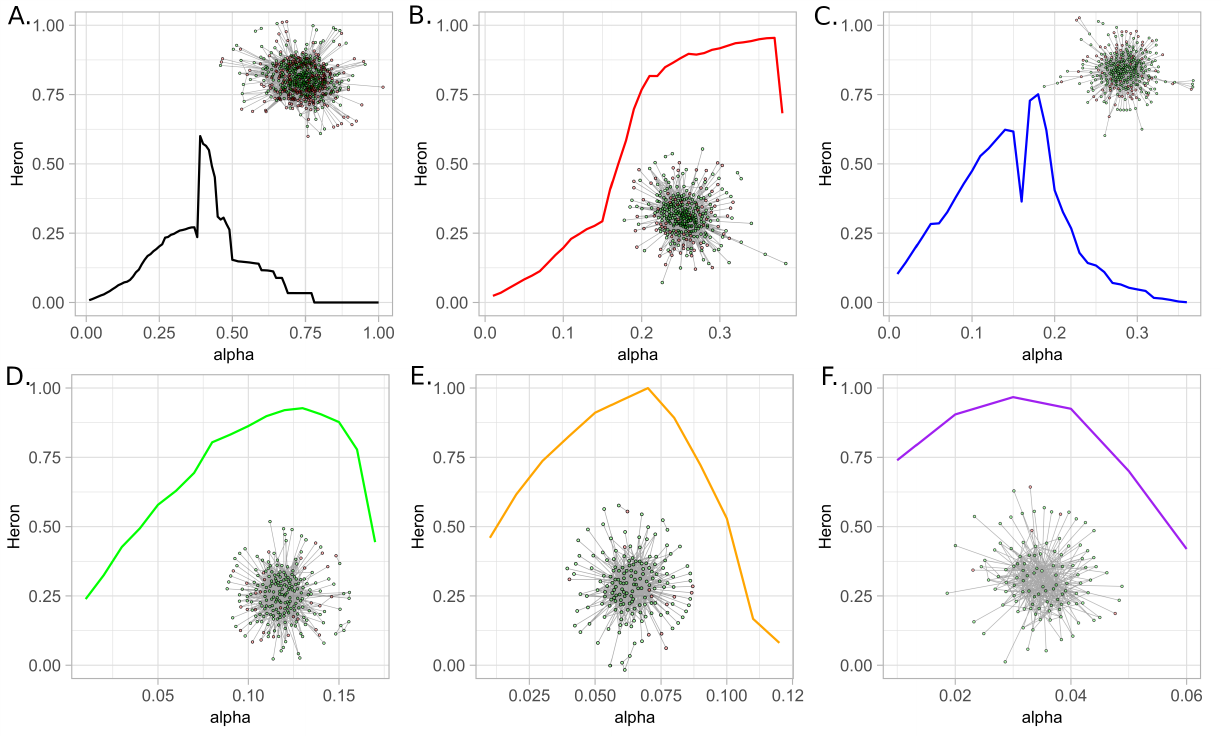}
 \caption{Significance level $\alpha$ over six iterations for the tube network in 2021, from A to F. The graph also shows the shrinking networks, as edges are removed every step with our method, where green nodes are companies that won at least one bidding and red nodes never won. }
 \label{fig:tubofinal}
\end{figure}

The most informative subgraph is chosen by maximizing the Heron Coefficient in the network, as illustrated in Figure \ref{fig:tubofinal}, which depicts the aggregate bidding processes for tubes in 2021. This optimization is performed iteratively, adjusting the significance level $\alpha$ at each step to maximize the network’s Heron Coefficient. Interestingly, the reduction of the significance level implies in an increase of the fraction of winners and, at the sixth step the resulting network usually contains only winners companies.

\begin{figure}[h!]
 \centering
 \includegraphics[width=\linewidth,keepaspectratio=true]{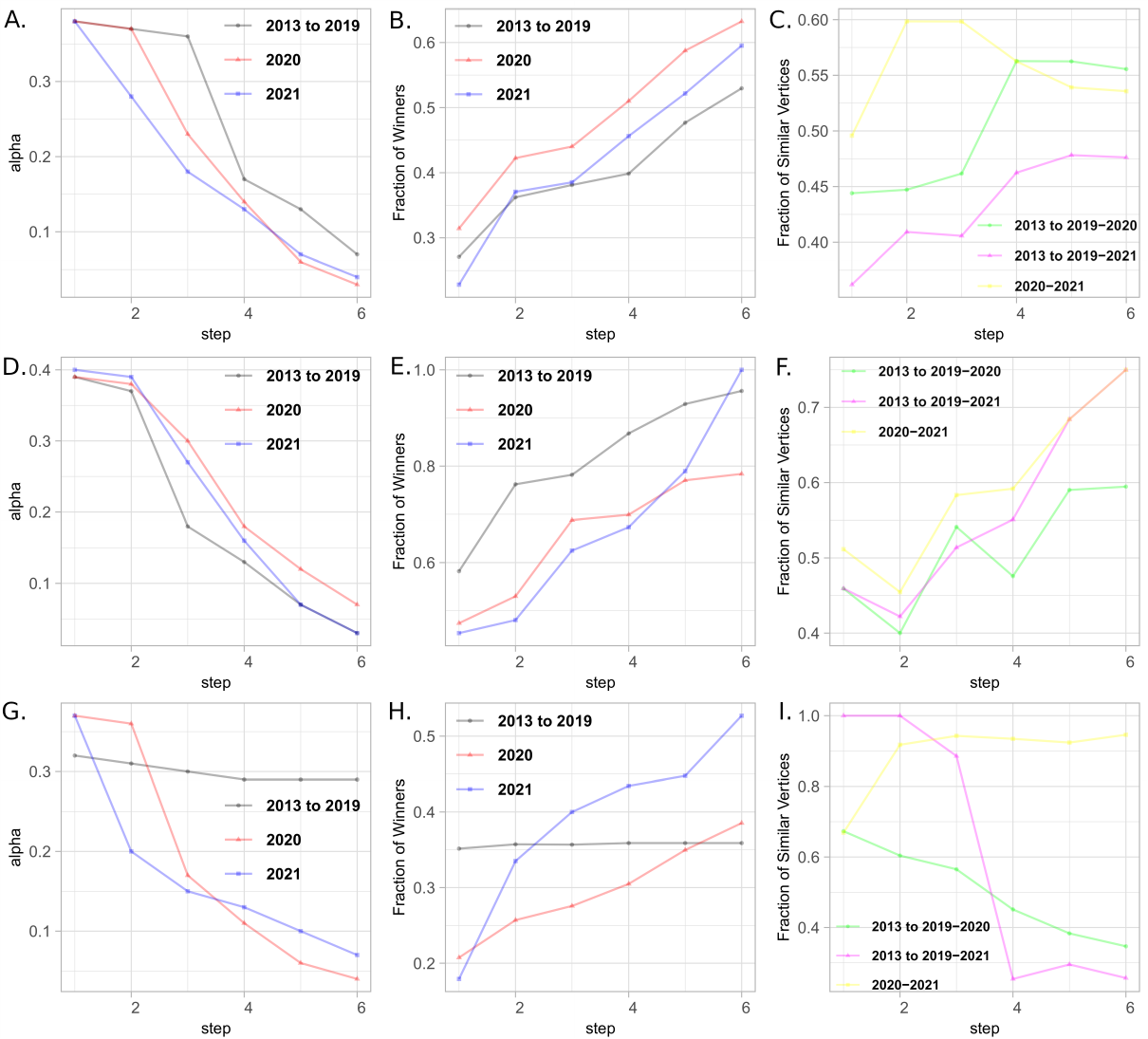}
 \caption{Evolution of the significance level ($\alpha$), fraction of winners and fraction of similar nodes for gloves (panels A to C), tubes (panels D to F) and masks (panels G to I) for each one of the algorithm's steps.}
 \label{fig:figuraevo}
\end{figure}

Figure \ref{fig:figuraevo} shows the evolution of the highest significance level of links in the networks, the percentage of winning companies in the network and the common elements in 3 different periods: pre-pandemic (2013-2019), the pandemic's first year (2020) and its second year (2021). Further details of the significance level's experiments and results can be checked on the Supplementary Material \ref{S1}.

The evolution of $\alpha$ with the removal of redundant links is similar for gloves (panels A to C) and endotracheal tubes (panels D to F), both before and after the pandemic. The six steps quickly find a subgraph with relevant links, with $\alpha \leq 10\%$. Concurrently, the proportion of winning companies within these subgraphs increases at each step, indicating that the most significant connections persist among successful bidders. In the tubes network, all companies in the sixth step had won at least one bid. The most significant difference between these two networks emerges in their temporal dynamics. The gloves network exhibits a stable composition, with 2021 closely resembling 2020 and earlier years (2013–2019), reflecting market continuity. In contrast, the tubes network underwent a marked disruption in 2020, followed by a reversion in 2021 toward its pre-2020 structure. Although many firms exited the market between 2019 and 2020, several long-standing participants reappeared in 2021. Thus, the data for 2013–2019 can predict 2021 participants, but not those from the anomalous year of 2020. This evolution illustrates the sensitivity of the Heron metric to shifts in market structure: its earlier peaks captured potential collusive patterns, while the 2020 surge signals renewed concentration that warrants closer monitoring. 

For a more in-depth analysis, we tested the Heron Information Coefficient shifts with sliding windows every trimester. By examining the HIC value at a point in time and contrasting it with the expected value given previous behavior, we further stress the temporal aspects of the evolving networks. Here, we can better differentiate the behavior of the gloves and tubes networks.

\begin{figure}[h]
  \centering
  \includegraphics[width=\linewidth]{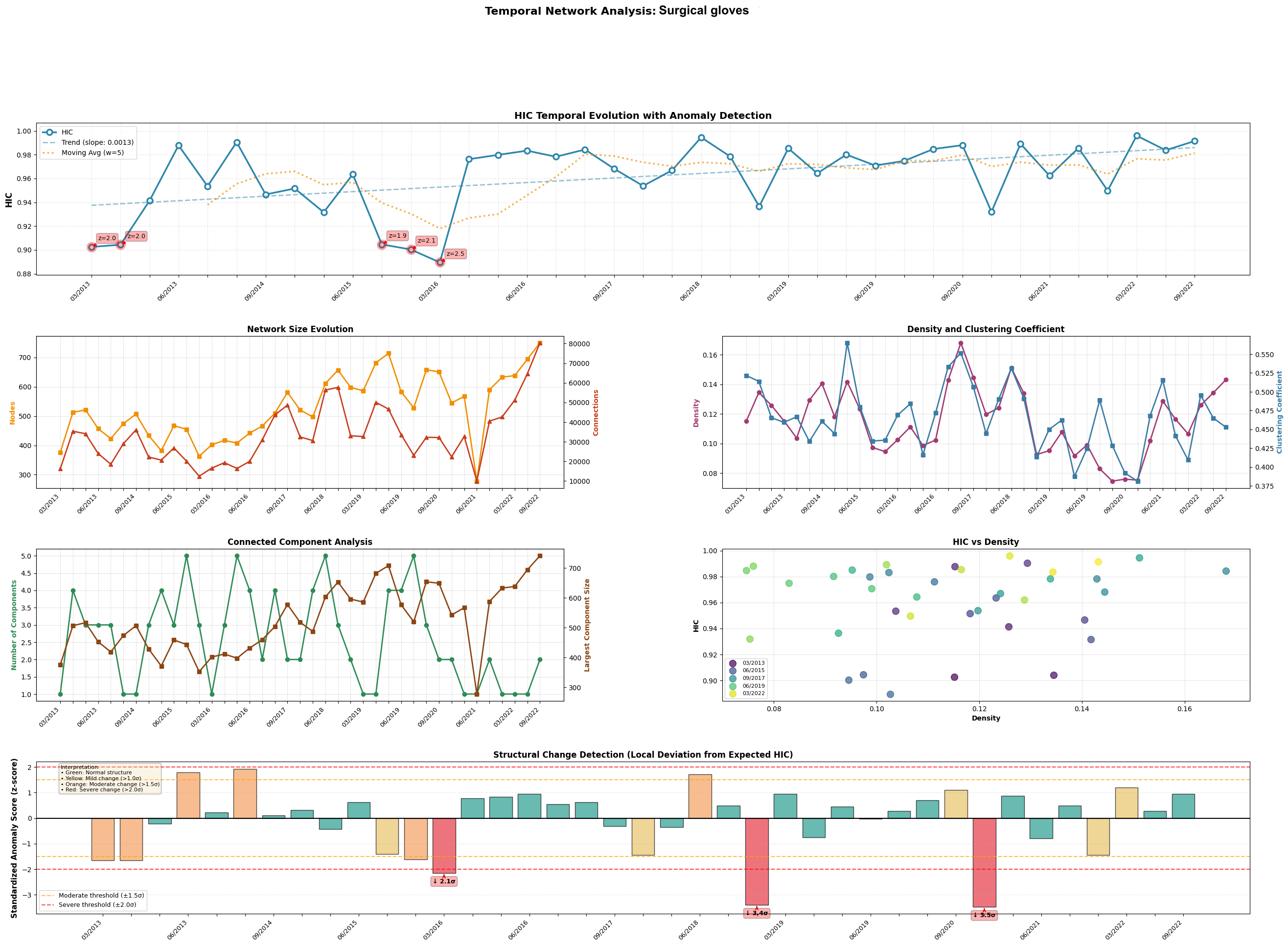}
  \caption{Summary of sliding windows experiment of the Surgical Glove network from 2013 to 2022, per trimesters. The pandemic had an expressive effect on the network behavior.}\label{luvatemp}
\end{figure}

In the surgical gloves network, shown in Figure \ref{luvatemp}, HIC values remain consistently high, but moderate anomalies emerge in 2019 and 2020, marking localized reorganizations rather than systemic disruptions during the pandemic. The smooth upward trend and moderate variability in density and clustering both suggest a gradual structural change, with increasing efficiency achieved through selective, rather than expansive, connectivity. The persistence of multiple components in earlier intervals, followed by consolidation, reinforces the interpretation of slow but steady and selective market concentration.

\begin{figure}[h]
  \centering
  \includegraphics[width=\linewidth]{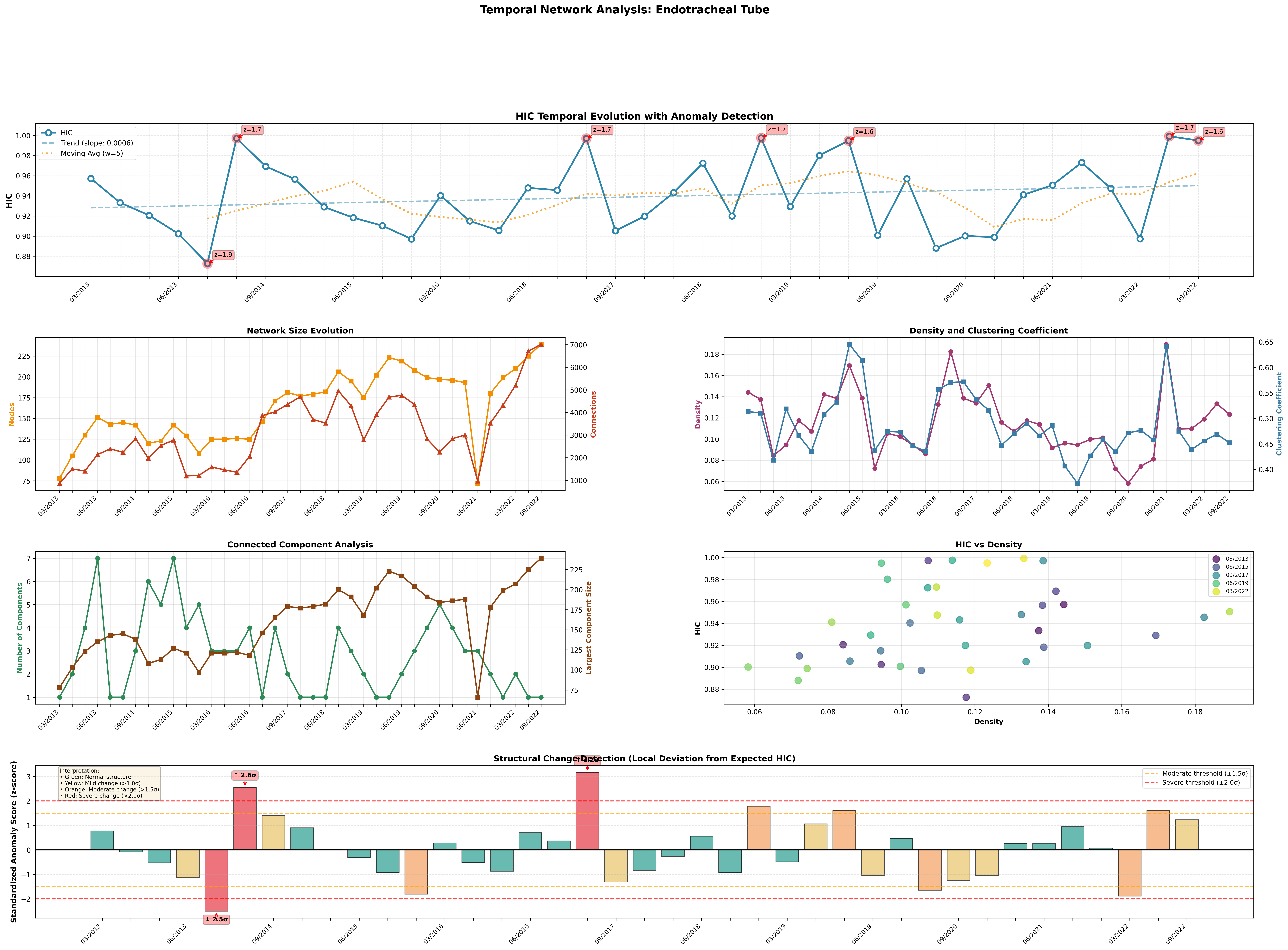}
  \caption{Summary of sliding windows experiment of the Endotracheal Tube network from 2013 to 2022, per trimesters. Its pre-pandemic behavior is quite different from the other networks analyzed. }\label{tubotemp}
\end{figure}

In the network of endotracheal tubes, seen in Figure \ref{tubotemp}, we find a distinct pattern of structural contraction. HIC remains stable at high values for most of the period, but decreases sharply toward the end, coinciding with a reduction in nodes and edges. The presence of strong positive and negative anomaly scores signals alternating phases of expansion and collapse, suggesting a system characterized by transient structural fragility rather than gradual stabilization. Therefore, even though the general trend is the same as the gloves network, its finer details have significant differences.

The network of masks was the most affected by the pandemic. In 2019, 928 companies were competing. In 2020, this number increased to 2558; in 2021, it decreased to 1673 - which is still an expressive increase compared to pre-pandemic levels. Unlike the other networks, the mask network underwent a clear phase transition as seen in Figure \ref{fig:figuraevo}. Before the pandemic, the $\alpha$ failed to isolate a subgraph of statistically significant links. In 2020 and 2021, however, the $\alpha$ is reduced, indicating that the remaining links were more structurally relevant. During these years, the proportion of winning companies increased at each filtering step, contrasting with the pre-pandemic pattern where this percentage remained nearly constant. After 2020, significant connections formed preferentially among winning firms, signaling a reorganization of the competitive structure. This drastic shift in the network's behavior and composition persisted into 2021, involving largely the same dominant players. 

\begin{figure}[h]
  \centering
  \includegraphics[width=\linewidth]{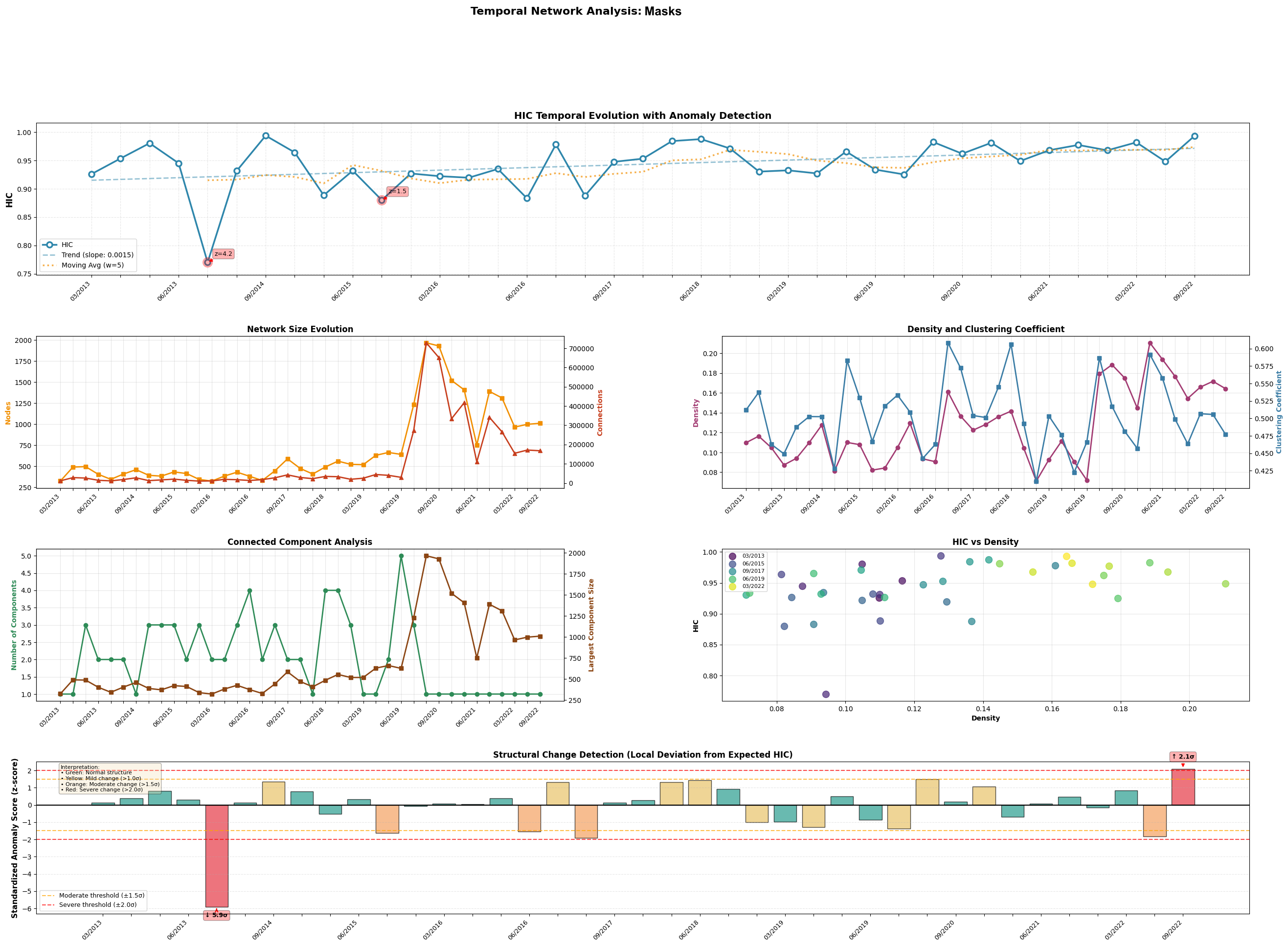}
  \caption{Summary of sliding windows experiment of the Surgical Mask network from 2013 to 2022, per trimesters. This network was particularly affected by the pandemic.}\label{masktemp}
\end{figure}

More specifically, as shown in Figure \ref{masktemp}, the masks network exhibits both large-scale expansion and pronounced anomalies, in contrast to the other two. A severe deviation in early 2020 coincides with a major increase in network size and connections, reflecting abrupt topological shifts as expected with the pandemic. The subsequent HIC recovery and stabilization indicate rapid adaptation and reorganization toward a more heavily dominated market configuration.

These experimental results show the effects on different networks before and after an unparalleled disturbance. For gloves, there are higher link importance levels in the subgraphs, with increasingly more winners and a gradual differentiation of the participating companies over the analyzed periods. There is a higher link's importance level for tubes but a less accurate selection of companies in the first pandemic year. Even then, 2021 has a higher percentage of winning companies than before the pandemic and similar elements to previous years after the six steps. Finally, there is a dramatic shift in significance level for masks, winning remaining companies and composition when comparing the pre-pandemic scenario with the pandemic. The pandemic has higher significance levels and percentages of winning companies, as well as similar arrangements in 2020 and 2021 that are very dissimilar to previous years. The sliding-window evolution of HIC effectively mirrors changes in network centrality distributions and topological robustness—high and stable HIC values correspond to configurations with strong central nodes and cohesive connectivity cores, whereas low or volatile HIC signals fragmentation and vulnerability to structural perturbations.

Overall, our method captures the significance of the remaining links and elements in the system. Allied with the composition comparison with previous years, this approach composes a detailed characterization of the system and its dynamics. Across all three cases, the inverse relationship between HIC and density remains consistent. Moreover, the method’s anomaly detection accurately pinpoints critical transition points, distinguishing the gradual evolution seen with surgical gloves and endotracheal tubes from more abrupt reconfigurations, as seen with masks. These results confirm the robustness and interpretability of our HIC-based temporal framework. Our method effectively captures both smooth and disruptive structural dynamics and identifies critical change intervals, providing a useful tool for covert networks detection.

This work establishes Heron’s Information Coefficient (HIC) as a robust metric for quantifying structural imbalances within dynamic networks. Applied to Brazilian health procurement data during the COVID-19 pandemic, HIC successfully isolated topological anomalies consistent with collusive behavior and market concentration—findings further corroborated by stress-testing on synthetic graphs with varying corruption densities. While the method's geometric intuition effectively exposes deviations often smoothed over by conventional statistics, we emphasize that it functions best as a complementary diagnostic rather than a standalone proof of illicit activity, particularly given the inherent scarcity of ground-truth criminal data. Future research should focus on benchmarking HIC against established information-theoretic metrics and developing real-time monitoring pipelines, thereby broadening its applicability to procurement integrity, fraud detection, and the resilience of complex systems.

\section{Conclusion}
This work introduces Heron’s Information Coefficient as a robust measure to quantify structural imbalances in dynamic networks. Applied to Brazilian health procurement data spanning the COVID-19 pandemic, HIC successfully isolated topological anomalies consistent with collusive behavior and market concentration - findings further corroborated by stress-testing on synthetic graphs with varying corruption densities and an aggregate procurement dataset. 

While promising, this approach should be treated as complementary to existing detection frameworks and not as a stand-alone diagnostic. The method's geometric intuition helps interpret deviations that conventional measures might smooth over, but it does not capture all institutional or legal nuances. As in many studies of covert criminal networks, our work also faces limited access to reliable real-world data for broader validation. We emphasize that our method functions best as a complementary diagnostic rather than a standalone proof of illicit activity, particularly given the inherent scarcity of ground-truth criminal data and the ever-evolving dynamics of criminal activity. 

Future studies should systematically compare the Heron coefficient with other dissimilarity or entropy-based metrics and explore its use in other anomaly detection scenarios. We believe that this measure can add value to research on procurement integrity, fraud prevention, the resilience of complex systems, as well as, more broadly, community detection.

\bibliographystyle{elsarticle-num} 
\bibliography{references}

@article{garciarodriguezCollusionDetectionPublic2022,
  title = {Collusion Detection in Public Procurement Auctions with Machine Learning Algorithms},
  author = {Garc{\'i}a Rodr{\'i}guez, Manuel J. and {Rodr{\'i}guez-Montequ{\'i}n}, Vicente and {Ballesteros-P{\'e}rez}, Pablo and Love, Peter E. D. and Signor, Regis},
  year = 2022,
  month = jan,
  journal = {Automation in Construction},
  volume = {133},
  pages = {104047},
  issn = {0926-5805},
  doi = {10.1016/j.autcon.2021.104047},
  urldate = {2026-01-12}
}

@book{Klette_Rosenfeld_2004, title={Digital Geometry}, DOI={10.1016/b978-1-55860-861-0.x5000-7}, journal={Elsevier eBooks}, author={Klette, Reinhard and Rosenfeld, Azriel}, year={2004}, month=jan, pages={77–116}, publisher={Elsevier} }

@article{akbarpour_diffusion_2018,
	title = {Diffusion in networks and the virtue of burstiness},
	volume = {115},
	issn = {0027-8424, 1091-6490},
	doi = {10.1073/pnas.1722089115},
	pages = {E6996--E7004},
	number = {30},
	journal = {Proceedings of the National Academy of Sciences},
	shortjournal = {{PNAS}},
	author = {Akbarpour, Mohammad and Jackson, Matthew O.},
	urldate = {2020-09-23},
	date = {2018-07-24},
    year = {2018},
	pmid = {29987048},
	note = {Publisher: National Academy of Sciences
Section: {PNAS} Plus}
}

@article{akogluGraphBasedAnomaly2015,
  title = {Graph Based Anomaly Detection and Description: A Survey},
  shorttitle = {Graph Based Anomaly Detection and Description},
  author = {Akoglu, Leman and Tong, Hanghang and Koutra, Danai},
  year = 2015,
  month = may,
  journal = {Data Mining and Knowledge Discovery},
  volume = {29},
  number = {3},
  pages = {626--688},
  issn = {1573-756X},
  doi = {10.1007/s10618-014-0365-y},
  urldate = {2026-01-12},
  langid = {english}
}

@article{maComprehensiveSurveyGraph2023,
  title = {A {{Comprehensive Survey}} on {{Graph Anomaly Detection With Deep Learning}}},
  author = {Ma, Xiaoxiao and Wu, Jia and Xue, Shan and Yang, Jian and Zhou, Chuan and Sheng, Quan Z. and Xiong, Hui and Akoglu, Leman},
  year = 2023,
  month = dec,
  journal = {IEEE Transactions on Knowledge and Data Engineering},
  volume = {35},
  number = {12},
  pages = {12012--12038},
  issn = {1558-2191},
  doi = {10.1109/TKDE.2021.3118815},
  urldate = {2026-01-12}
}

@article{serranoExtractingMultiscaleBackbone2009,
  title = {Extracting the Multiscale Backbone of Complex Weighted Networks},
  author = {Serrano, M. {\'A}ngeles and Bogu{\~n}{\'a}, Mari{\'a}n and Vespignani, Alessandro},
  year = {2009},
  journal = {Proceedings of the National Academy of Sciences},
  volume = {106},
  number = {16},
  pages = {6483--6488},
  doi = {10.1073/pnas.0808904106}
}

@article{zhouInfluenceInterlinkTopology2020,
  title = {Influence of {{Interlink Topology}} on {{Multilayer Network Robustness}}},
  author = {Zhou, Fang and He, Xiang and Yuan, Yongbo and Zhang, Mingyuan},
  year = {2020},
  month = jan,
  journal = {Sustainability},
  volume = {12},
  number = {3},
  pages = {1202},
  publisher = {Multidisciplinary Digital Publishing Institute},
  doi = {10.3390/su12031202},
  urldate = {2020-07-03},
  copyright = {http://creativecommons.org/licenses/by/3.0/},
  langid = {english}
}

@article{ficaraCovertNetworkConstruction2022,
  title = {Covert {{Network Construction}}, {{Disruption}}, and {{Resilience}}: {{A Survey}}},
  shorttitle = {Covert {{Network Construction}}, {{Disruption}}, and {{Resilience}}},
  author = {Ficara, Annamaria and Curreri, Francesco and Fiumara, Giacomo and De Meo, Pasquale and Liotta, Antonio},
  year = {2022},
  month = jan,
  journal = {Mathematics},
  volume = {10},
  number = {16},
  pages = {2929},
  publisher = {Multidisciplinary Digital Publishing Institute},
  issn = {2227-7390},
  doi = {10.3390/math10162929},
  urldate = {2025-10-13},
  copyright = {http://creativecommons.org/licenses/by/3.0/},
  langid = {english}
}

@article{campanaExplainingCriminalNetworks2016,
  title = {Explaining Criminal Networks: {{Strategies}} and Potential Pitfalls},
  shorttitle = {Explaining Criminal Networks},
  author = {Campana, Paolo},
  year = {2016},
  month = jan,
  journal = {Methodological Innovations},
  volume = {9},
  pages = {2059799115622748},
  publisher = {SAGE Publications Ltd},
  issn = {2059-7991},
  doi = {10.1177/2059799115622748},
  urldate = {2025-10-06}
}

@article{cavallaroDisruptingResilientCriminal2020,
  title = {Disrupting Resilient Criminal Networks through Data Analysis: {{The}} Case of {{Sicilian Mafia}}},
  shorttitle = {Disrupting Resilient Criminal Networks through Data Analysis},
  author = {Cavallaro, Lucia and Ficara, Annamaria and Meo, Pasquale De and Fiumara, Giacomo and Catanese, Salvatore and Bagdasar, Ovidiu and Song, Wei and Liotta, Antonio},
  year = {5 de ago. de 2020},
  journal = {PLOS ONE},
  volume = {15},
  number = {8},
  pages = {e0236476},
  publisher = {Public Library of Science},
  issn = {1932-6203},
  doi = {10.1371/journal.pone.0236476},
  urldate = {2025-10-06},
  langid = {english}
}

@article{aytac_network_2018,
	title = {Network robustness and residual closeness},
	volume = {52},
	issn = {0399-0559, 1290-3868},
	doi = {10.1051/ro/2016071},
	pages = {839--847},
	number = {3},
	journal = {{RAIRO} - Operations Research},
	author = {Aytaç, A. and Berberler, Z. N. O.},
	urldate = {2020-07-03},
	year = {2018}
}

@article{bachmann_survey_2020,
	title = {A Survey on Frameworks Used for Robustness Analysis on Interdependent Networks},
	type = {Review Article},
	author = {Bachmann, Ivana and Bustos-Jiménez, Javier and Bustos, Benjamin},
	year = {2020},
	journal={Complexity}, publisher={Hindawi},
	doi = {10.1155/2020/2363514},
	ISSN={1076-2787}
}

@article{otsuka_robustness_2019,
	title = {Robustness of network attack strategies against node sampling and link errors},
	volume = {14},
	issn = {1932-6203},
	doi = {10.1371/journal.pone.0221885},
	pages = {e0221885},
	number = {9},
	journal = {{PLOS} {ONE}},
	author = {Otsuka, Momoko and Tsugawa, Sho},
	urldate = {2020-07-03},
	year = {2019}
}

@article{wu_enhancing_2017,
	title = {Enhancing structural robustness of scale-free networks by information disturbance},
	volume = {7},
	issn = {2045-2322},
	doi = {10.1038/s41598-017-07878-2},
	pages = {7559},
	number = {1},
	journal = {Scientific Reports},
	author = {Wu, Jun and Tan, Suo-Yi and Liu, Zhong and Tan, Yue-Jin and Lu, Xin},
	urldate = {2020-07-02},
	year = {2017}
}

@article{tejedor_network_2017,
	title = {Network robustness assessed within a dual connectivity framework: joint dynamics of the Active and Idle Networks},
	volume = {7},
	issn = {2045-2322},
	doi = {10.1038/s41598-017-08714-3},
	pages = {8567},
	number = {1},
	journal = {Scientific Reports},
	author = {Tejedor, Alejandro and Longjas, Anthony and Zaliapin, Ilya and Ambroj, Samuel and Foufoula-Georgiou, Efi},
	urldate = {2020-07-02},
	year = {2017}
}

@article{albert_error_2000,
	title = {Error and attack tolerance of complex networks},
	volume = {406},
	issn = {1476-4687},
	doi = {10.1038/35019019},
	pages = {378--382},
	number = {6794},
	journal = {Nature},
	author = {Albert, R. and Jeong, H. and Barabási, A.-L.},
	urldate = {2020-05-12},
	year = {2000}
}

@article{schieber_quantification_2017,
	title = {Quantification of network structural dissimilarities},
	volume = {8},
	issn = {2041-1723},
	doi = {10.1038/ncomms13928},
	pages = {1--10},
	number = {1},
	journal = {Nature Communications},
	author = {Schieber, T. A. and  Laura Carpi and Albert Díaz-Guilera and Panos M. Pardalos and Cristina Masoller and Martín G. Ravetti},
	urldate = {2020-03-31},
	year = {2017}
}

@article{wang_entropy_2006,
	title = {Entropy optimization of scale-free networks’ robustness to random failures},
	volume = {363},
	issn = {0378-4371},
	doi = {10.1016/j.physa.2005.08.025},
	pages = {591--596},
	number = {2},
	journal = {Physica A: Statistical Mechanics and its Applications},
	author = {Wang, B. and Huanwen Tang and Chonghui Guo and Zhilong Xiu},
	urldate = {2020-03-30},
	year = {2006}
}

@article{smith_competitive_2019,
	title = {Competitive percolation strategies for network recovery},
	volume = {9},
	issn = {2045-2322},
	doi = {10.1038/s41598-019-48036-0},
	pages = {1--12},
	number = {1},
	journal = {Scientific Reports},
	author = {Smith, A. M.  and  Márton Pósfai and Martin Rohden and Andrés D. González and Leonardo Dueñas-Osorio and Raissa M. D’Souza },
	urldate = {2020-03-30},
	year = {2019}
}

@article{collins_decentralising_2000,
	title = {Decentralising the health sector: issues in {Brazil}},
	volume = {52},
	issn = {0168-8510},
	DOI = {10.1016/S0168-8510(00)00069-5},
	pages = {113--127},
	number = {2},
	journal = {Health Policy},
	author = {Collins, Charles and Araujo, Jose and Barbosa, Jarbas},
	urldate = {2020-10-13},
	year = {2000}
}

@article{wachs_network_2019,
	title = {A network approach to cartel detection in public auction markets},
	volume = {9},
	issn = {2045-2322},
	doi = {10.1038/s41598-019-47198-1},
	pages = {10818},
	number = {1},
	journal = {Scientific Reports},
	author = {Wachs, Johannes and Kertész, János},
	urldate = {2020-10-13},
	year = {2019},
	note = {Number: 1, Publisher: Nature Publishing Group}
}

@Article{lin1991,
  author      = {Lin, Jianhua},
  title       = {Divergence measures based on the Shannon entropy},
  journal     = {IEEE Transactions on Information Theory},
  year        = {1991},
  volume      = {37},
  pages       = {145-151},
  issn        = {0018-9448},
  doi         = {10.1109/18.61115}
}

@article{puppim_de_oliveira_brazilian_2017,
	title = {Brazilian Public Administration: Shaping and Being Shaped by Governance and Development},
	volume = {2},
	issn = {2365-4252},
	doi = {10.1007/s41111-017-0052-4},
	shorttitle = {Brazilian Public Administration},
	pages = {7--21},
	number = {1},
	journal = {Chinese Political Science Review},
	author = {Puppim de Oliveira, Jose A.},
	urldate = {2021-03-06},
	year = {2017}
}

@article{ErdosRenyi1960, 
    title={On the evolution of random graphs}, author={Erdős, P. and Renyi, A.},
    journal = {Bull. Inst. Internat. Statist},
    volume = {38},
    issue = {4},
    year={1960} 
}

@misc{noauthor_ibge_nodate,
	title = {Projeção da população},
        author = {IBGE},
	note = {\url{https://www.ibge.gov.br/apps/populacao/projecao/index.html}},
	year = {2021}
}

@article{paim_brazilian_2011,
	title = {The Brazilian health system: history, advances, and challenges},
	volume = {377},
	issn = {01406736},
	doi = {10.1016/S0140-6736(11)60054-8},
	pages = {1778--1797},
	number = {9779},
	journal = {The Lancet},
	author = {Paim, Jairnilson and Travassos, Claudia and Almeida, Celia and Bahia, Ligia and Macinko, James},
	year = {2011}
}

@techreport{harrington_detecting_2005,
	title = {Detecting Cartels},
	url = {https://econpapers.repec.org/paper/jhupapers/526.htm},
	institution = {The Johns Hopkins University,Department of Economics},
	type = {Economics Working Paper Archive},
	author = {Harrington, Joseph},
	urldate = {2021-03-13},
	year = {2005}
}

@misc{noauthor_exploitative_2020,
    author = {OECD},
	title = {Exploitative pricing in the time of {COVID}},
	url = {https://www.oecd.org/daf/competition/Exploitative-pricing-in-the-time-of-COVID-19.pdf},
	year = {2020}
}

@article{harrington_cartel_2006,
	title = {Cartel Pricing Dynamics with Cost Variability and Endogenous Buyer Detection},
	doi = {10.1016/j.ijindorg.2006.04.012},
	pages = {1185--1212},
	journal = {International Journal of Industrial Organization},
	author = {Harrington, Joseph and Chen, Joe},
	year = {2006}
}

@article{kohler_does_2015,
	title = {Does Pharmaceutical Pricing Transparency Matter? Examining Brazil’s Public Procurement System},
	volume = {11},
	issn = {1744-8603},
	doi = {10.1186/s12992-015-0118-8},
	shorttitle = {Does Pharmaceutical Pricing Transparency Matter?},
	pages = {34},
	number = {1},
	journal = {Globalization and Health},
	shortjournal = {Global Health},
	author = {Kohler, Jillian Clare and Mitsakakis, Nicholas and Saadat, Faridah and Byng, Danalyn and Martinez, Martha Gabriela},
	urldate = {2021-03-14},
	year = {2015}
}

@article{teremetskyi_corruption_2021,
	title = {Corruption and strengthening anti-corruption efforts in healthcare during the pandemic of Covid-19},
	volume = {89},
	issn = {0025-8172},
	doi = {10.1177/0025817220971925},
	pages = {25--28},
	number = {1},
	journal = {Medico-Legal Journal},
	author = {Teremetskyi, Vladyslav and Duliba, Yevheniia and Kroitor, Volodymyr and Korchak, Nataliia and Makarenko, Oleksandr},
	urldate = {2021-03-14},
	year = {2021},
	note = {Publisher: {SAGE} Publications}
}

@article{vian_review_2008,
	title = {Review of corruption in the health sector: theory, methods and interventions},
	volume = {23},
	issn = {0268-1080},
	doi = {10.1093/heapol/czm048},
	pages = {83--94},
	number = {2},
	journal = {Health Policy and Planning},
	author = {Vian, Taryn},
	urldate = {2021-03-14},
	year = {2008}
}

@article{kirya_anti-corruption_nodate,
	title = {Anti-corruption in Covid-19 preparedness and response},
	journal={Bergen: U4 Anti-Corruption Resource Centre, Chr. Michelsen Institute. (U4 Brief 2020:8)},
	author = {Kirya, Monica},
	year={2020}
}

@article{wesseljun_22_its_2020,
	title = {{‘It’s a nightmare.’ How Brazilian} scientists became ensnared in chloroquine politics},
	author = {Wessel, L.},
	urldate = {2021-03-14},
	year = {2020},
	journal={Science Magazine},
        doi = {10.1126/science.abd4620}
}

@misc{darlington_how_nodate,
	title = {How {Brazil} gambled on unproven drugs to fight Covid-19},
	author = {Jose Brito Darlington},
	date = {2021-03-14},
	year = {2021},
	note={CNN.}
}

@misc{minas_exercito_2021,
	title = {{Exército e Ministério da Saúde} gastaram milhões para distribuir cloroquina},
        author = {Matheus Adler},
	urldate = {2021-03-14},
	year = {2021},
	note = { Estado de Minas }
}

@misc{steingruber_corruption_nodate,
	title = {Corruption in the time of {COVID}-19: A double-threat for low income countries},
	pages = {18},
	author = {Steingrüber, Sarah and Kirya, Monica and Jackson, David and Mullard, Saul},
	note={Bergen: U4 Anti-Corruption Resource Centre, Chr. Michelsen Institute (U4 Brief 2020:6)},
    year={2020}
}

@book{gragnolati_twenty_2013,
	title = {Twenty Years of Health System Reform in Brazil: An Assessment of the Sistema Único de Saúde},
	isbn = {978-0-8213-9843-2 978-0-8213-9932-3},
	shorttitle = {Twenty Years of Health System Reform in Brazil},
	publisher = {The World Bank},
	author = {Gragnolati, Michele and Lindelow, Magnus and Couttolenc, Bernard},
	urldate = {2021-03-14},
	year = {2013},
	doi = {10.1596/978-0-8213-9843-2}
}

@article{de_almeida_botega_brazilian_2020,
	title = {Brazilian hospitals’ performance: an assessment of the unified health system ({SUS})},
	volume = {23},
	issn = {1572-9389},
	doi = {10.1007/s10729-020-09505-5},
	pages = {443--452},
	number = {3},
	journal = {Health Care Management Science},
	shortjournal = {Health Care Manag Sci},
	author = {de Almeida Botega, Laura and Andrade, Mônica Viegas and Guedes, Gilvan Ramalho},
	urldate = {2021-03-14},
	year = {2020}
}

@article{almeida_health_2000,
	title = {Health Sector Reform in Brazil: A Case Study of Inequity},
	volume = {30},
	issn = {0020-7314},
	doi = {10.2190/NDGW-C2DP-GNF8-HEW8},
	pages = {129--162},
	number = {1},
	journal = {International Journal of Health Services},
	shortjournal = {Int J Health Serv},
	author = {Almeida, Celia and Travassos, Claudia and Porto, Silvia and Labra, Maria Eliana},
	urldate = {2021-03-14},
	year = {2000},
	note = {Publisher: {SAGE} Publications Inc}
}

@incollection{viana_universalizing_2017,
	location = {London},
	title = {{Universalizing Health Care in Brazil}: Opportunities and Challenges},
	isbn = {978-1-137-53377-7},
	booktitle = {Social Policy in a Development Context},
	shorttitle = {Universalizing Health Care in Brazil},
	pages = {181--211},
	publisher = {Palgrave Macmillan {UK}},
	author = {Viana, Ana Luiza d’Ávila and da Silva, Hudson Pacífico and Yi, Ilcheong},
	urldate = {2021-03-15},
	year = {2017},
	doi = {10.1057/978-1-137-53377-7_7},
}

@incollection{mattos_procurement_2016,
	title = {Procurement Procedures and Bid-Rigging in Brazil},
	isbn = {978-3-319-30948-4},
	series = {International Law and Economics},
	pages = {169--185},
	booktitle = {Competition Law Enforcement in the {BRICS} and in Developing Countries: Legal and Economic Aspects},
	publisher = {Springer International Publishing},
	author = {Mattos, César},
	urldate = {2021-03-15},
	year = {2016},
	DOI = {10.1007/978-3-319-30948-4_6}
}

@misc{controladoria-geral_da_uniao_licitacoes_nodate,
	title = {{Licitações} realizadas - {Portal} da transparência},
        author = {CGU},
	shorttitle = {Licitações realizadas},
	year = {2021},
	note = {\url{https://www.transparencia.gov.br/download-de-dados/licitacoes}}
}

@article{elsisyNetworkGeneratorCovert2022,
  title = {A Network Generator for Covert Network Structures},
  author = {Elsisy, Amr and Mandviwalla, Aamir and Szymanski, Boleslaw K. and Sharkey, Thomas},
  year = {2022},
  month = jan,
  journal = {Information Sciences},
  volume = {584},
  pages = {387--398},
  issn = {0020-0255},
  doi = {10.1016/j.ins.2021.10.066},
  urldate = {2025-01-28}
}

\appendix

\newpage
\setcounter{section}{0}
\setcounter{page}{1}
\setcounter{figure}{0}
\setcounter{equation}{0}
\renewcommand{\thesection}{S\arabic{section}}
\renewcommand{\thepage}{s\arabic{page}}
\renewcommand{\thetable}{S\arabic{table}}
\renewcommand{\thefigure}{S\arabic{figure}}

\setcounter{equation}{0}
\setcounter{figure}{0}
\setcounter{table}{0}
\setcounter{page}{1}
\makeatletter
\renewcommand{\theequation}{S\arabic{equation}}
\renewcommand{\thefigure}{S\arabic{figure}}
\renewcommand{\bibnumfmt}[1]{[S#1]}
\setcounter{affn}{0}
\resetTitleCounters

\makeatletter
\let\@title\@empty
\makeatother

\title{Supplementary Data for: Structural Asymmetry as a Fraud Signature: Detecting Collusion with Heron’s Information Coefficient}

\def\ps@pprintTitle{%
     \let\@oddhead\@empty
     \let\@evenhead\@empty
     \def\@oddfoot{\footnotesize\itshape
        Supplementary Data for \ifx\@journal\@empty Elsevier
       \else\@journal\fi\hfill\today}%
     \let\@evenfoot\@oddfoot}
\makeatother

\maketitle
\section{Complexity analysis}

In its standard formulation, the algorithm exhibits a time complexity of $\mathcal{O}(|V|(|V|+|E|))$ and a space complexity of $\mathcal{O}(|V|^2)$. While these bounds are manageable for small-to-medium networks, they present significant scaling challenges for graphs exceeding $10^4$ nodes, primarily due to memory constraints. The performance cost is driven by the \textit{Network Node Dispersion} (NND) function. Computing the topological dispersion requires an All-Pairs Shortest Path (APSP) calculation. For unweighted graphs, this is implemented via Breadth-First Search (BFS) traversals from each node, yielding a time complexity of $\mathcal{O}(|V|(|V|+|E|))$. More critically, the naive storage of the resulting distance matrix necessitates $\mathcal{O}(|V|^2)$ space, creating a dense memory bottleneck even for sparse input graphs.

This cost is compounded by the iterative Heron optimization loop. To identify the optimal cut, the algorithm sweeps through $K$ quantile steps (typically $K=100$), requiring dual NND evaluations at each step. If the pruning process continues for $M$ generations, the total theoretical complexity scales to $\mathcal{O}(M \cdot K \cdot |V| \cdot |E|)$.

A critical design choice in our framework is the exclusion of the alpha-centrality term ($w_3=0$) for large-scale analysis. Including this term requires operations on the graph complement - transforming sparse inputs into dense representations - which would elevate the time complexity to $\mathcal{O}(|V|^3)$. While the third term enhances sensitivity to spectral isomorphism, its exclusion is a necessary and acceptable approximation for backbone extraction. In this context, preserving the dominant connectivity patterns (captured by the first two terms) takes precedence over the detection of subtle spectral symmetries, rendering the cubic computational cost unjustifiable.

To overcome the $\mathcal{O}(|V|^2)$ memory barrier, we implemented a batched parallelization strategy. Because the graph topology evolves sequentially during pruning, caching is ineffective. Instead, we utilize:

\begin{enumerate}
    \item \textbf{Batched Pathfinding:} Rather than allocating a monolithic distance matrix, we compute shortest paths in strip-wise batches. Statistics (entropy and histograms) are aggregated on-the-fly, and raw distance data is immediately discarded. This reduces memory usage to $\mathcal{O}(|V| \times \text{batch\_size})$.
    \item \textbf{Integrated Diameter Calculation:} The network diameter is derived dynamically during the pathfinding batch, eliminating the need for a separate $\mathcal{O}(|V||E|)$ traversal.
    \item \textbf{Parallel Execution:} The batching process allows for efficient distribution across multi-core architectures, providing near-linear speedup in the NND calculation phase.
\end{enumerate}

This optimized approach shifts the bottleneck from memory capacity to CPU throughput, making the method viable for larger systems without sacrificing the rigorous topological analysis required by the disparity filter.

\section{Heron comparison in real data}\label{S1}

The way the vertices interact and connect provides essential information about the system. In particular, information about their connectivity patterns represents a valuable tool on how different dynamics occur in the system. For example, the speed at which information travels within a cohesive community is much greater than in a more sparse one. The theoretical benchmark of public bidding network consists of a network where nodes are connected uniformly. For example, in a total of 100 biddings, if a company $x$ participates in 10 ($n_x=10$) and a company $y$ participates of 15 ($n_y=15$), it is expected that the probability that these two companies participate in the same bidding is $0.1\times 0.15$, resulting in an average of 1.5 joint participation. Values above 1.5 must indicate some abnormal relationships. Here, we identify the evolution of links happening in a random uniform manner and links that occur with a higher probability than the expected null hypothesis {\cite{akbarpour_diffusion_2018}}.

Supposing a Poisson behavior, the probability of that element interacting with another depends only on the node itself as we count its frequency of participation. However, the probability of the elements participating in the same bidding when there is collusion differs from this uniform connectivity, affecting the system's topology.

To capture how different a random network composed of Poisson-like elements is from the real interaction between the elements, we built a $G_P$ network (Poisson network) from the fact that the probability of a link existing between two companies depends only on the number of times they participate in a contest. This network is compared with the real $G_R$ network where the probability of a link occurring is given by the fraction of its weight ($w$) by the total number of bids. More specifically, let $G=(V,E,W_v,W_e)$ be the bidding network of an item where $V$ is the set of vertices, $E$ the set of links, $W_v$ the set of the number of bids that each vertex participated in and $W_e$ the number of times the link $e$ occurred, $G_P=(V,E)$ the unweighted, where the probability of a link occurring is Poisson and $G_R$ the real network. We then compare the Heron's values for these two different networks generations. Table \ref{tabela} presents the evolution of $H_{real}$ versus the expected $H_{null}$. We report the ratio $H_{real}/H_{null}$ to illustrate the magnitude of structural deviation. Bold values indicate statistical significance ($p < 0.05$), determined via a Z-test against the null ensemble distribution. A ratio significantly greater than 1.0 implies that the information geometry of the real network is fundamentally more complex than what can be explained by simple participation rates, warranting further investigation by competent authorities.

\begin{table}[h!] \label{tabela}
\centering
\caption{The Heron coefficients' annual evolution for the Real network ($G_R$), considering a Poisson behavior ($G_P$), the fraction between the Heron coefficients (Real/Poisson), and the total spent in Brazilian Real for each product bid. Bold values of the fraction between the coefficients mean that, on average, the Heron and Poisson coefficients are different at a 5\% significance level.}
\begin{adjustbox}{width=\linewidth}
\begin{tabular}{lccccccccc}
\toprule
\textbf{Year} & 
\multicolumn{3}{c}{\textbf{Hospital Apron}} & 
\multicolumn{3}{c}{\textbf{Surgical Mask}} & 
\multicolumn{3}{c}{\textbf{Surgical Glove}} \\
\cmidrule(lr){2-4} \cmidrule(lr){5-7} \cmidrule(lr){8-10}
 & Heron (Real) & Heron (Poisson) & Fraction 
 & Heron (Real) & Heron (Poisson) & Fraction 
 & Heron (Real) & Heron (Poisson) & Fraction \\
\midrule
2013 & 0.746 & 0.274 & \textbf{2.720 } 
     & 0.008 & 0.002 & \textbf{3.625}  
     & 0.080 & 0.115 & \textbf{0.697}  \\

2014 & 0.515 & 0.209 & \textbf{2.470}  
     & 0.008 & 0.002 & \textbf{4.105}  
     & 0.110 & 0.221 & \textbf{0.521}  \\

2015 & 0.686 & 0.275 & \textbf{2.498} 
     & 0.009 & 0.008 & 1.056  
     & 0.102 & 0.241 & 0.838 \\

2016 & 0.189 & 0.040 & \textbf{4.693} 
     & 0.020 & 0.005 & \textbf{5.475} 
     & 0.122 & 0.257 & \textbf{0.475} \\

2017 & 0.079 & 0.081 & 0.980  
     & 0.050 & 0.013 & \textbf{3.759}  
     & 0.188 & 0.419 & \textbf{0.449}  \\

2018 & 0.069 & 0.051 & 1.338  
     & 0.050 & 0.013 & \textbf{1.997} 
     & 0.155 & 0.363 & \textbf{0.426}  \\

2019 & 0.035 & 0.040 & 0.881  
     & 0.010 & 0.003 & \textbf{2.942} 
     & 0.099 & 0.297 & \textbf{0.333}  \\

2020 & 0.059 & 0.024 & \textbf{2.513}  
     & 0.018 & 0.007 & \textbf{2.369}  
     & 0.086 & 0.226 & \textbf{0.380}  \\
\bottomrule
\end{tabular}
\end{adjustbox}
\end{table}

\section{The BID network evolution} \label{S2}

In this section, we show the networks of three items: surgical masks, surgical gloves, and hospital gowns. These graphs offer a more in-depth visualization of the networks analyzed in the sliding windows experiment.

\begin{figure}[h]
  \centering
  \includegraphics[width=\linewidth]{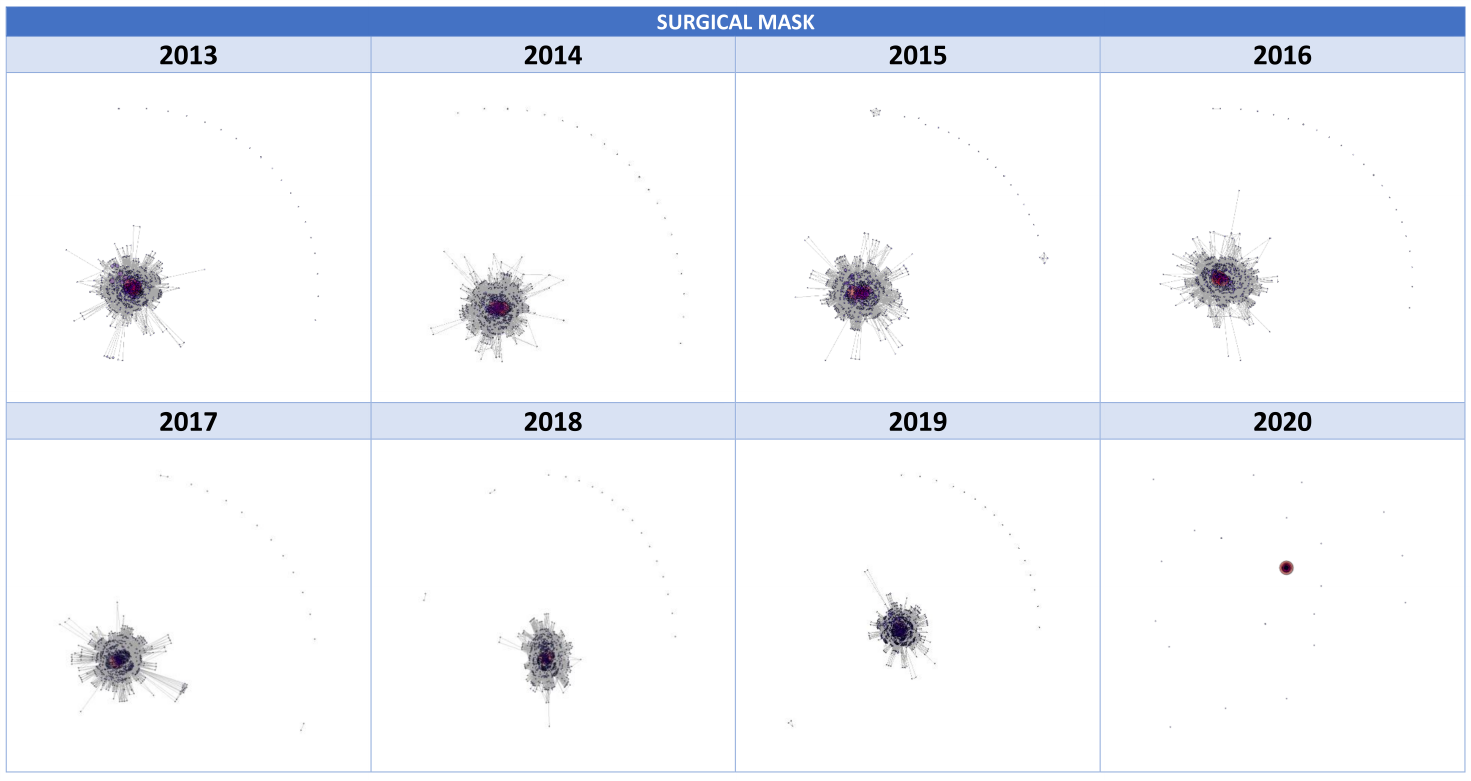}
  \caption{Evolution of the  Surgical Mask network from 2013 to 2020. The amount of bidding process that a company participates in sizes the vertices.}\label{mascara_cir}
\end{figure}

Figure \ref{mascara_cir} shows the networks formed with surgical masks' bidding processes. With the exception of 2015, the real network's Heron coefficient was systematically higher than the aleatory network's. As such, this network has a consistent behavior in important companies. In 2020, with the COVID-19 pandemic, the higher demand was accompanied by an expressive increase of bidding companies. Even so, the difference between the Heron's coefficients in this case shows the persistence of previous behavior in some links that should be further investigated by the competent authorities.

\begin{figure}[h]
  \centering
  \includegraphics[width=\linewidth]{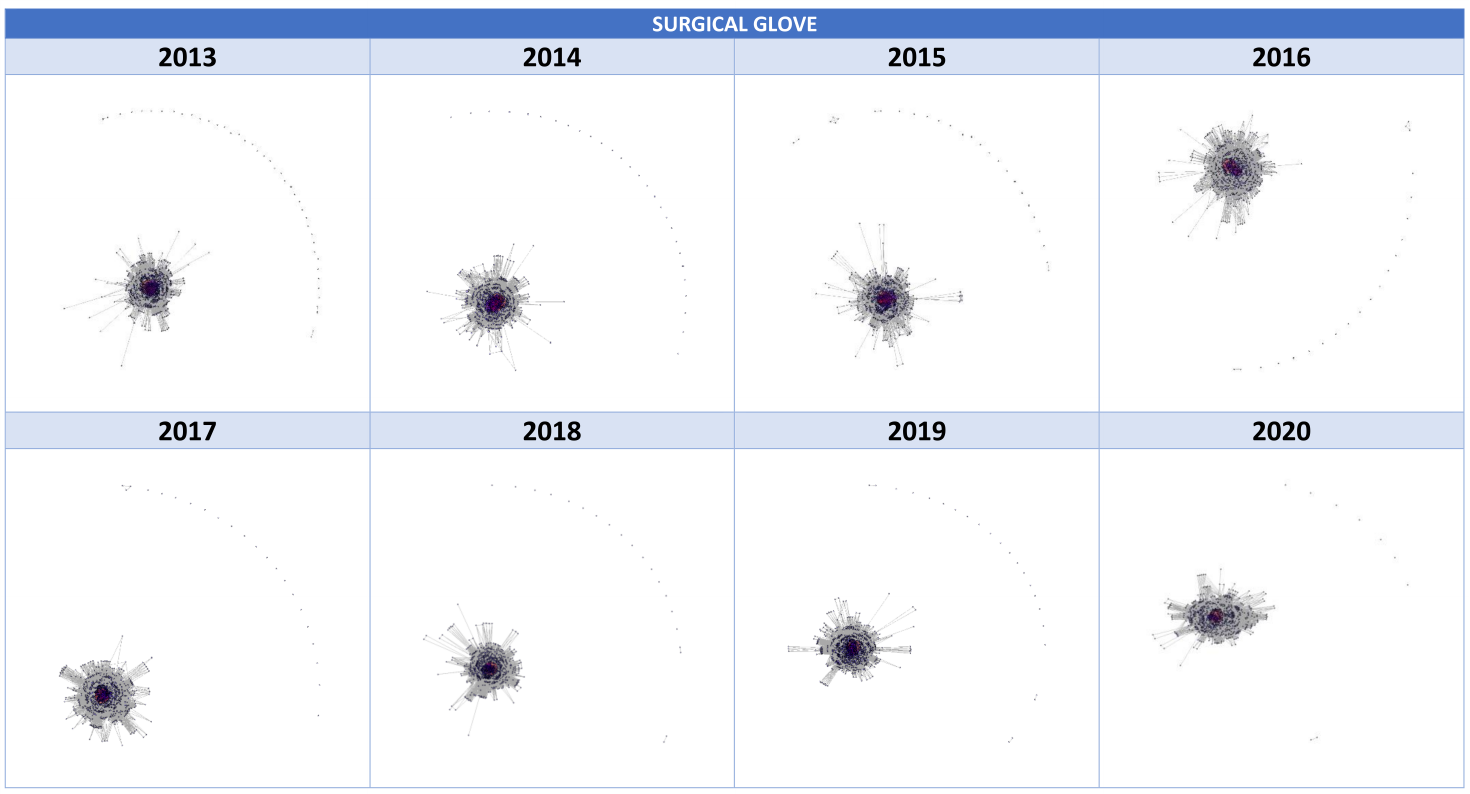}
  \caption{Evolution of the Surgical Glove network from 2013 to 2020. The amount of bidding process that a company participates in sizes the vertices.}\label{luva_cir}
\end{figure}

Figure \ref{luva_cir} illustrates the surgical gloves' bidding network. As we can see, from 2013 to 2020, there were few changes in network's topology, with a more homogeneous topology. In this case, the aleatory network's Heron's coefficient is often significantly higher than the real network's. Even then, there is a suspicious concentration at the end of 2020 that should be further investigated by the competent authorities. 

\begin{figure}[h]
  \centering
  \includegraphics[width=\linewidth]{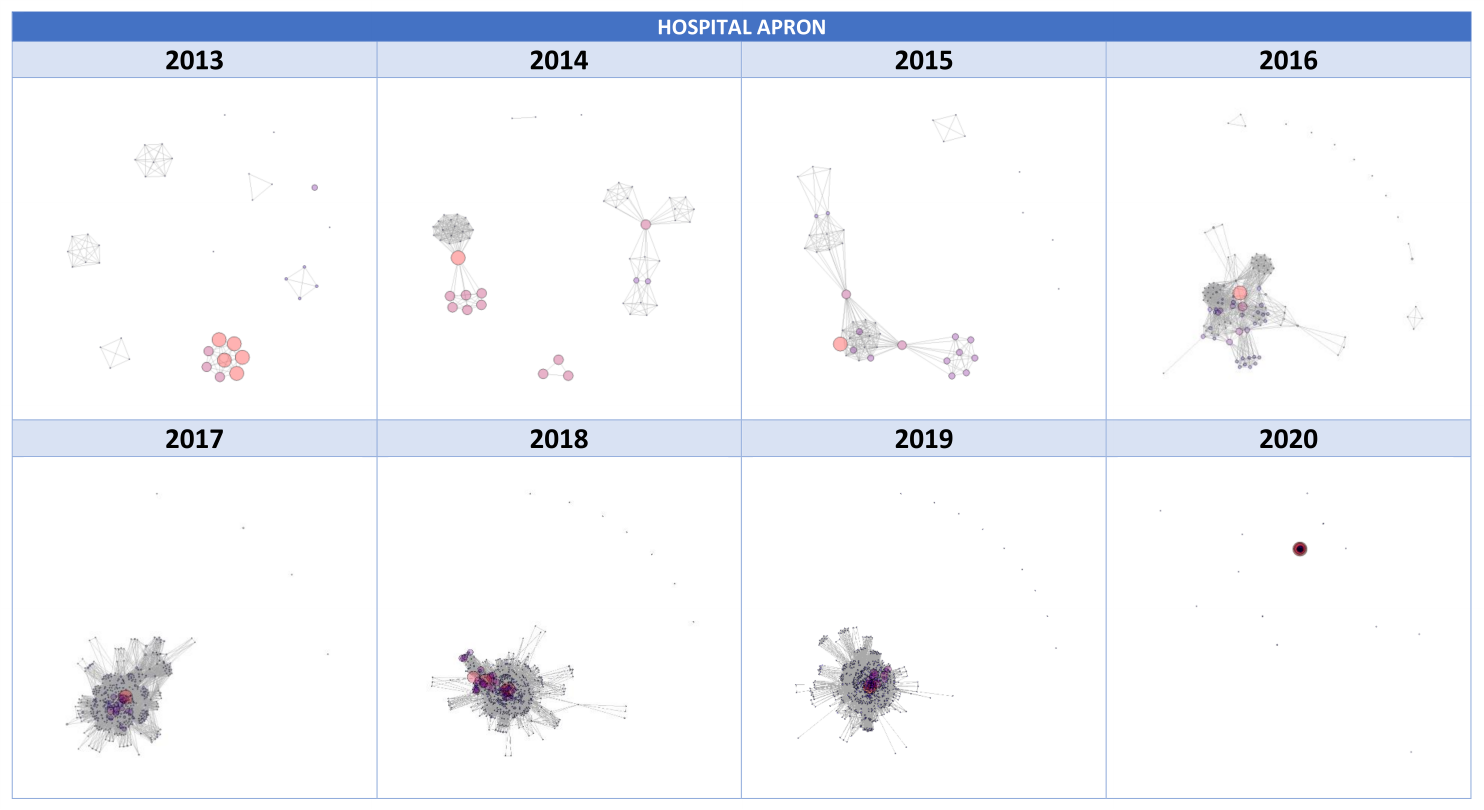}
  \caption{Evolution of the Hospital Surgical Gowns network from 2013 to 2020. The amount of bidding process that a company participates in sizes the vertices.}\label{hospital_apron}
\end{figure}

\begin{figure}[h]
  \centering
  \includegraphics[width=\linewidth]{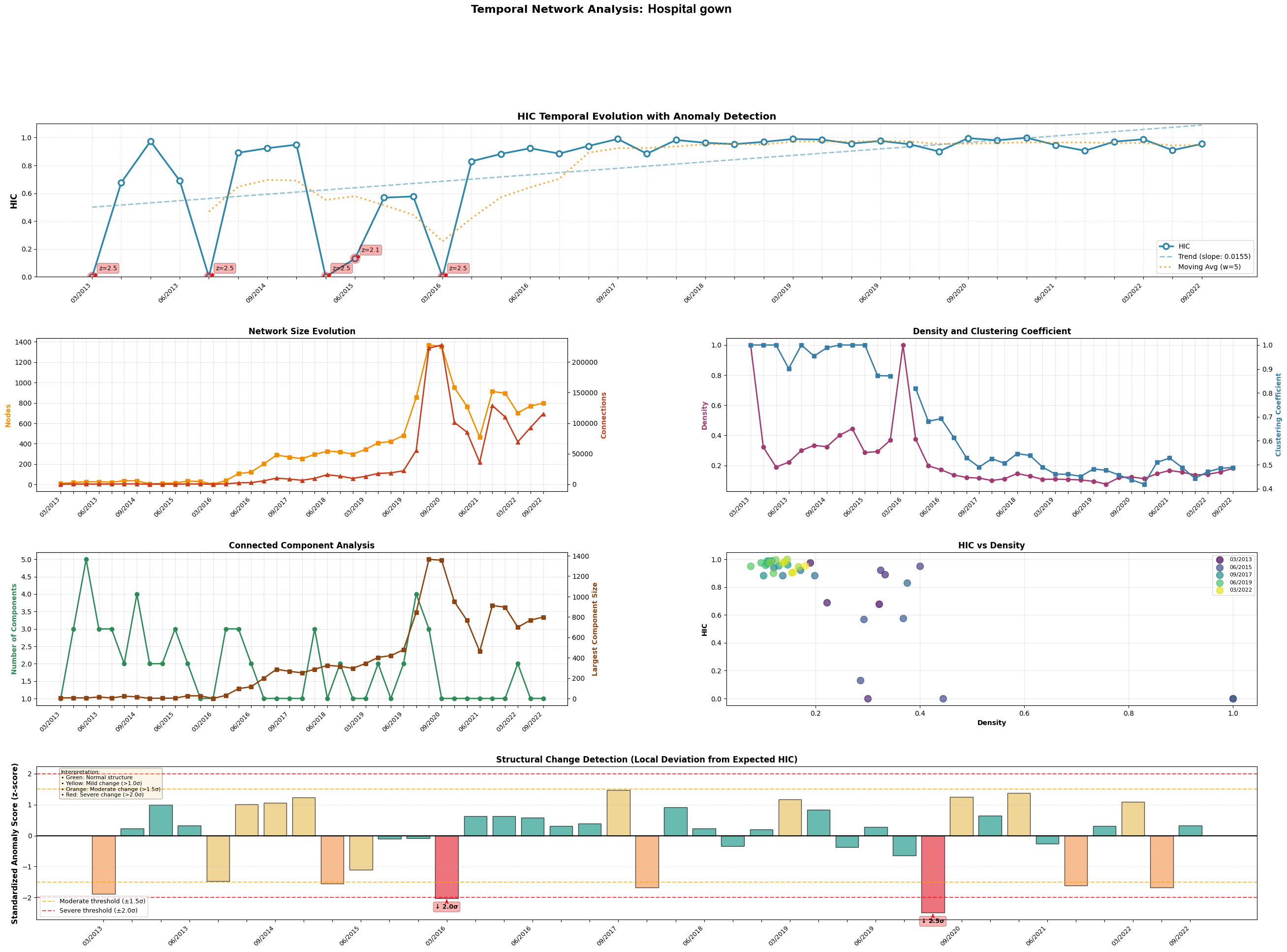}
  \caption{Summary of sliding windows experiment of the Hospital Apron network from 2013 to 2022, per trimesters. Its pre-pandemic behavior is quite different from the other networks analyzed. }\label{hosptemp}
\end{figure}

Figures \ref{hospital_apron} and \ref{hosptemp} depict the bidding network for another item: hospital aprons. The evolution in the hospital aprons' bidding process can be taken as a great visual example of the changes captured by the Heron's coefficient, as it is the network that changed the most over time. In 2013, the network exhibited a low structural complexity, characterized by a few tightly connected hubs. Its Heron coefficient was substantially higher than the expected aleatory value, suggesting non-random concentration of bids. The same dominant companies remained central until 2015, with minor structural variations yet consistently elevated Heron values. The ratio between the real and aleatory coefficients increased further, peaking in 2016. Between 2017 and 2019, the entry of new participants enhanced competitiveness, driving the network closer to random behavior. In 2020, however, a few firms regained dominance, reflected in another marked rise in the Heron coefficient. These structural changes were properly captured by HIC, with two major deviations in early 2019 and mid-2020, where standardized anomaly scores exceed the significance threshold. This evolution illustrates the sensitivity of the Heron metric to shifts in market structure: its earlier peaks captured potential collusive patterns, while the 2020 surge signals renewed concentration that warrants closer monitoring.

\end{document}